\RequirePackage[2020/02/02]{latexrelease}
\documentclass[
  nojss]{jss}

\usepackage[utf8]{inputenc}

\author{
Derek Beaton\\Rotman Research Institute, Baycrest Health Sciences
}
\title{Generalized eigen, singular value, and partial least squares
decompositions: The \pkg{GSVD} package}

\Plainauthor{Derek Beaton}
\Plaintitle{Generalized eigen, singular value, and partial least squares
decompositions: The GSVD package}
\Shorttitle{\pkg{GSVD}: generalized SVD}

\Abstract{
The generalized singular value decomposition (GSVD, a.k.a. ``SVD
triplet'', ``duality diagram'' approach) provides a unified strategy and
basis to perform nearly all of the most common multivariate analyses
(e.g., principal components, correspondence analysis, multidimensional
scaling, canonical correlation, partial least squares). Though the GSVD
is ubiquitous, powerful, and flexible, it has very few implementations.
Here I introduce the \pkg{GSVD} package for \proglang{R}. The general
goal of \pkg{GSVD} is to provide a small set of accessible functions to
perform the GSVD and two other related decompositions (generalized
eigenvalue decomposition, generalized partial least squares-singular
value decomposition). Furthermore, \pkg{GSVD} helps provide a more
unified conceptual approach and nomenclature to many techniques. I first
introduce the concept of the GSVD, followed by a formal definition of
the generalized decompositions. Next I provide some key decisions made
during development, and then a number of examples of how to use
\pkg{GSVD} to implement various statistical techniques. These examples
also illustrate one of the goals of \pkg{GSVD}: how others can (or
should) build analysis packages that depend on \pkg{GSVD}. Finally, I
discuss the possible future of \pkg{GSVD}.
}

\Plainkeywords{multivariate analysis, generalized singular value decomposition, principal components analysis, correspondence analysis, multidimensional scaling, canonical correlation, partial least squares, R}

\Address{
    Derek Beaton\\
    Rotman Research Institute, Baycrest Health Sciences\\
    3560 Bathurst Street Toronto, ON Canada M6A 2E1\\
  E-mail: \email{dbeaton@research.baycrest.org}\\
  
  }

\usepackage{booktabs}
\usepackage{longtable}
\usepackage{array}
\usepackage{multirow}
\usepackage{wrapfig}
\usepackage{float}
\usepackage{colortbl}
\usepackage{pdflscape}
\usepackage{tabu}
\usepackage{threeparttable}
\usepackage{threeparttablex}
\usepackage[normalem]{ulem}
\usepackage{makecell}
\usepackage{xcolor}

\usepackage{amsmath} \usepackage{booktabs} \usepackage{float} \usepackage{amssymb}

\begin{document}

\hypertarget{introduction}{%
\section{Introduction}\label{introduction}}

The singular value decomposition \citep[SVD;][]{golub_singular_1971} is
one of the most important tools in a multivariate toolbox. Conceptually
and practically, the SVD is the core technique behind numerous
statistical methods; the most common of which is principal component
analysis
\citep[PCA;][]{jolliffe_principal_2012, abdi_principal_2010, jolliffe_principal_2016}.
A lesser known---but far more ubiquitous---tool is the generalized SVD
\citep[GSVD;][]{abdi2007singular, greenacre_theory_1984, takane_relationships_2003, holmes_multivariate_2008}.
The GSVD is more ubiquitous because it is the technique behind---and
generalizes---many more statistical methods. The core concept behind the
GSVD is that a data matrix has two companion matrices: one for the rows
and one for the columns. These companion matrices are weights,
constraints, or metrics imposed on the rows and columns of a data
matrix. For examples, via the GSVD we can implement weighted solutions
to PCA, or analyses under different metrics, such as correspondence
analysis
\citep[CA;][]{greenacre_theory_1984, lebart_multivariate_1984, escofier-cordier_analyse_1965, greenacre_correspondence_2010}
or canonical correlation analysis
\citep[CCA;][]{harold1936relations, abdi2017canonical}.

Though the GSVD generalizes---and is more flexible than---the SVD, there
are few if any direct implementations of the GSVD in \proglang{R}.
Rather, the GSVD is typically part of specific packages, and these
packages are typically designed around more general analyses and broad
usage. Rarely, if ever, are these GSVD implementations accessible to, or
meant for, more direct access by users. Given the frequent and
ubiquitous uses of the SVD and GSVD, a more accessible implementation
would benefit a wide range of users.

Here I introduce a package designed around the GSVD, called \pkg{GSVD}.
\pkg{GSVD} is a lightweight implementation of the GSVD and two other
generalized decompositions: the generalized eigendecomposition (GEVD)
and the generalized partial least squares-SVD (GPLSSVD). \pkg{GSVD} was
designed for a wide range of users, from analysts to package developers,
all of whom would benefit from more direct access to the GSVD and
similar decompositions. More importantly, the \pkg{GSVD} package and the
idea of the GSVD provide a basis to unify concepts, nomenclature, and
techniques across a wide array of statistical traditions and
multivariate analyses approaches. \pkg{GSVD} has three core functions:
\code{geigen()}, \code{gsvd()}, and \code{gplsssvd()}. These core
functions provide a way for users to implement a wide array of methods
including (but not limited to) multidimensional scaling, principal
components analysis, correspondence analysis, canonical correlation,
partial least squares, and numerous variants and extensions of the
aforementioned. \pkg{GSVD} also helps simplify and unify concepts across
techniques because, at their core, all of these techniques can be
accomplished with the SVD.

In this paper, I introduce the core functions and functionality of
\pkg{GSVD}. Furthermore, I show how the GSVD can provide a more unified
nomenclature across techniques, and I show how various techniques can be
implemented through \code{geigen()}, \code{gsvd()}, and
\code{gplsssvd()}. Here I provide sufficient detail for the majority of
users and readers. However, some readers may want to take a deeper dive
into various techniques and literatures. Where possible and relevant, I
point the readers to works with much more detail and substance.

This paper is outlined as follows. In \emph{Generalized decompositions}
I provide background on, notation for, and mathematical explanations of
the three decompositions discussed here (GEVD, GSVD, and GPLSSVD),
followed by some additional notes and literature. In \emph{Package
description}, I provide an overview of the core functions, their uses,
and other development notes. In \emph{Examples of multivariate analyses}
I provide detailed implementations of numerous multivariate techniques,
as well as variations of those techniques or variations of how to
implement them with \pkg{GSVD} (e.g., PCA via GEVD vs PCA via GSVD). In
\emph{Discussion} I make some comments about the package, its uses, its
potential, and possible future development.

\hypertarget{generalized-decompositions}{%
\section{Generalized decompositions}\label{generalized-decompositions}}

The GSVD is probably best known by way of the correspondence analysis
literature
\citep{greenacre_theory_1984, lebart_multivariate_1984, escofier-cordier_analyse_1965, greenacre_correspondence_2010}
and, more generally, the ``French way'' of multivariate data analyses
\citep{holmes_discussion_2017, holmes_multivariate_2008}. There is
considerable breadth and depth of materials on the GSVD and related
techniques since the 1960s (and even well before then). Though such
breadth and depth is beyond the scope of this paper, I provide a list of
references and resources throughout the paper for any interested
readers. This section instead focuses on the formulation---and to a
degree the nomenclature---of the three core techniques. I first
introduce notation, followed by the GEVD, the GSVD, and then formalize
and introduce the GPLSSVD.

\hypertarget{notation}{%
\subsection{Notation}\label{notation}}

Bold uppercase letters denote matrices (e.g., \(\mathbf{X}\)). Upper
case italic letters (e.g., \(I\)) denote cardinality, size, or length.
Subscripts for matrices denote relationships with certain other
matrices, for examples \({\mathbf Z}_{\mathbf X}\) is some matrix
derived from or related to the \({\bf X}\) matrix, where something like
\({\bf F}_{I}\) is a matrix related to the \(I\) set of elements. When
these matrices are introduced, they are also specified. Two matrices
side-by-side denotes standard matrix multiplication (e.g.,
\(\bf{X}\bf{Y}\)). Superscript \(^{T}\) denotes the transpose operation,
and superscript \(^{-1}\) denotes standard matrix inversion. The
diagonal operator, denoted \(\mathrm{diag\{\}}\), transforms a vector
into a diagonal matrix, or extracts the diagonal of a matrix in order to
produce a vector.

\hypertarget{generalized-eigendecomposition}{%
\subsection{Generalized
eigendecomposition}\label{generalized-eigendecomposition}}

The generalized eigendecomposition (GEVD) requires two matrices: a
\(J \times J\) (square) data matrix \({\bf X}\) and a \(J \times J\)
constraints matrix \({\bf W}_{\bf X}\). For the GEVD, \({\bf X}\) is
typically positive semi-definite and symmetric (e.g., a covariance
matrix) and \({\bf W}_{\bf X}\) is required to be positive
semi-definite. The GEVD decomposes the data matrix \({\bf X}\)---with
respect to its constraints \({\bf W}_{\bf X}\)---into two matrices as
\begin{equation}
{\bf X} = {\bf Q}{\bf \Lambda}{\bf Q}^{T},
\end{equation} where \({\bf \Lambda}\) is a diagonal matrix that
contains the eigenvalues and \({\bf Q}\) are the \emph{generalized}
eigenvectors. The GEVD finds orthogonal slices of a data matrix, with
respect to its constraints, where each orthogonal slice explains the
maximum possible variance. That is, the GEVD maximizes
\({\bf \Lambda} = {\bf Q}^{T}{\bf W}_{\bf X}{\bf X}{\bf W}_{\bf X}{\bf Q}\)
under the constraint of orthogonality where
\({\bf Q}^{T}{\bf W}_{\bf X}{\bf Q} = {\bf I}\). Practically, the GEVD
is performed with the standard eigenvalue decomposition (EVD) as
\begin{equation}
\widetilde{\bf X} = {\bf V}{\bf \Lambda}{\bf V},
\end{equation} where
\(\widetilde{\bf X} = {\bf W}_{\bf X}^{\frac{1}{2}}{\bf X}{\bf W}_{\bf X}^{\frac{1}{2}}\)
and \({\bf V}\) are the eigenvectors which are orthonormal, such that
\({\bf V}^{T}{\bf V} = {\bf I}\). The relationship between the GEVD and
EVD can be explained as the relationship between the generalized and
standard eigenvectors where \begin{equation}
{\bf Q} = {\bf W}_{\bf X}^{-\frac{1}{2}}{\bf V} \Longleftrightarrow {\bf V} = {\bf W}_{\bf X}^{\frac{1}{2}}{\bf Q}.
\end{equation} When \({\bf W}_{\bf X} = {\bf I}\), the GEVD produces
exactly the same results as the EVD because
\(\widetilde{\bf X} = {\bf X}\) and thus \({\bf Q} = {\bf V}\). Analyses
with the EVD and GEVD---such as PCA---typically produce component or
factor scores. With the GEVD, component scores are defined as
\begin{equation}
{\bf F}_{J} = {\bf W}_{\bf X}{\bf Q}{\bf \Delta},
\end{equation} where \({\bf \Delta} = {\bf \Lambda}^{\frac{1}{2}}\),
which are singular values. The maximization in the GEVD can be reframed
as the maximization of the component scores where
\({\bf \Lambda} = {\bf F}_{J}^{T}{\bf W}_{\bf X}^{-1}{\bf F}_{J}\),
still subject to \({\bf Q}^{T}{\bf W}_{\bf X}{\bf Q} = {\bf I}\).

\hypertarget{generalized-singular-value-decomposition}{%
\subsection{Generalized singular value
decomposition}\label{generalized-singular-value-decomposition}}

The generalized singular value decomposition (GSVD) requires three
matrices: an \(I \times J\) (rectangular) data matrix \({\bf X}\), an
\(I \times I\) row constraints matrix \({\bf M}_{\bf X}\), and a
\(J \times J\) columns constraints matrix \({\bf W}_{\bf X}\). For the
GSVD \({\bf M}_{\bf X}\) and \({\bf W}_{\bf X}\) are each required to be
positive semi-definite. The GSVD decomposes the data matrix
\({\bf X}\)---with respect to both of its constraints
\({\bf M}_{\bf X}\) and \({\bf W}_{\bf X}\)---into three matrices as

\begin{equation}
{\bf X} = {\bf P}{\bf \Delta}{\bf Q}^{T},
\end{equation} where \({\bf \Delta}\) is a diagonal matrix that contains
the singular values, and where \({\bf P}\) and \({\bf Q}\) are the left
and right \emph{generalized} singular vectors, respectively. From the
GSVD we can obtain eigenvalues as \({\bf \Lambda}^{2} = {\bf \Delta}\).
The GSVD finds orthogonal slices of a data matrix, with respect to its
constraints, where each slice explains the maximum possible \emph{square
root} of the variance. That is, the GSVD maximizes
\({\bf \Delta} = {\bf P}^{T}{\bf M}_{\bf X}{\bf X}{\bf W}_{\bf X}{\bf Q}\)
under the constraint of orthogonality where
\({\bf P}^{T}{\bf M}_{\bf X}{\bf P} = {\bf I} = {\bf Q}^{T}{\bf W}_{\bf X}{\bf Q}\).
Typically, the GSVD is performed with the standard SVD as
\begin{equation}
\widetilde{\bf X} = {\bf U}{\bf \Delta}{\bf V},
\end{equation} where
\(\widetilde{\bf X} = {{\bf M}_{\bf X}^{\frac{1}{2}}}{\bf X}{{\bf W}^{\frac{1}{2}}_{\bf X}}\),
and where \({\bf U}\) and \({\bf V}\) are the left and right singular
vectors, respectively, which are orthonormal such that
\({\bf U}^{T}{\bf U} = {\bf I} = {\bf V}^{T}{\bf V}\). The relationship
between the GSVD and SVD can be explained as the relationship between
the generalized and standard singular vectors where \begin{equation}
\begin{aligned}
{\bf P} = {{\bf M}^{-\frac{1}{2}}_{\bf X}}{\bf U} \Longleftrightarrow {\bf U} = {{\bf M}^{\frac{1}{2}}_{\bf X}}{\bf P} \\
{\bf Q} = {{\bf W}^{-\frac{1}{2}}_{\bf X}}{\bf V} \Longleftrightarrow {\bf V} = {{\bf W}^{\frac{1}{2}}_{\bf X}}{\bf Q}.
\end{aligned}
\end{equation} When \({\bf M}_{\bf X} = {\bf I} = {\bf W}_{\bf X}\), the
GSVD produces exactly the same results as the SVD because
\(\widetilde{\bf X} = {\bf X}\) and thus \({\bf P} = {\bf U}\) and
\({\bf Q} = {\bf V}\). Analyses with the SVD and GSVD---such as PCA or
CA---typically produce component or factor scores. With the GSVD,
component scores are defined as \begin{equation}
{\bf F}_{I} = {\bf M}_{\bf X}{\bf P}{\bf \Delta} \textrm{ and } {\bf F}_{J} = {\bf W}_{\bf X}{\bf Q}{\bf \Delta},
\end{equation} for the left (rows) and right (columns) of \({\bf X}\),
respectively. The optimization in the GSVD can be reframed as the
maximization of the component scores where
\({\bf F}_{I}^{T}{\bf M}_{\bf X}^{-1}{\bf F}_{I} = {\bf \Lambda} = {\bf F}_{J}^{T}{\bf W}_{\bf X}^{-1}{\bf F}_{J}\),
still subject to
\({\bf P}^{T}{\bf M}_{\bf X}{\bf P} = {\bf I} = {\bf Q}^{T}{\bf W}_{\bf X}{\bf Q}\).
Note how the optimization with respect to the component scores shows a
maximization for the eigenvalues.

\hypertarget{generalized-partial-least-squares-singular-value-decomposition}{%
\subsection{Generalized partial least squares singular value
decomposition}\label{generalized-partial-least-squares-singular-value-decomposition}}

The generalized partial least squares-singular value decomposition
(GPLSSVD) is a reformulation of the PLSSVD. The PLSSVD is a specific
type of PLS from the broader PLS family \citep{tenenhaus1998regression}.
The PLSSVD has various other names---for example, PLS correlation
\citep{krishnan_partial_2011}---but canonically comes from the
psychology \citep{ketterlinus1989partial} and neuroimaging literatures
\citep{mcintosh_spatial_1996}, though it traces its origins to Tucker's
interbattery factor analysis \citep{tucker_inter-battery_1958}.
Recently, with other colleagues, I introduced the idea of the GPLSSVD as
it helped us formalize a new version of PLS for categorical data
\citep{beaton_partial_2016, beaton_generalization_2019}. However, the
GPLSSVD also allows us to generalize other methods, such as canonical
correlation analysis \citep[see Supplemental Material
of][]{beaton_generalization_2019}.

The GPLSSVD requires six matrices: an \(N \times I\) (rectangular) data
matrix \({\bf X}\) with its \(N \times N\) row constraints matrix
\({\bf M}_{\bf X}\) and its \(I \times I\) columns constraints matrix
\({\bf W}_{\bf X}\), and an \(N \times J\) (rectangular) data matrix
\({\bf Y}\) with its \(N \times N\) row constraints matrix
\({\bf M}_{\bf Y}\) and its \(J \times J\) columns constraints matrix
\({\bf W}_{\bf Y}\). For the GPLSSVD all constraint matrices are
required to be positive semi-definite. The GPLSSVD decomposes \emph{the
relationship between} the data matrices, with respect to their
constraints, and expresses the common information as the relationship
between latent variables. The goal of partial least squares-SVD (PLSSVD)
is to find a combination of orthogonal latent variables that maximize
the relationship between two data matrices. PLS is often presented as
\(\mathrm{arg max(} {\bf {l_{\bf X}}_{\ell}^{T}}{{\bf l}_{\bf Y}}_{\ell}\mathrm{)} = \mathrm{arg max}\textrm{ }\mathrm{cov(} {\bf {l_{\bf X}}_{\ell}}, {{\bf l}_{\bf Y}}_{\ell}\mathrm{)}\),
under the condition that
\({\bf {l_{\bf X}}_{\ell}^{T}}{{\bf l}_{\bf Y}}_{\ell'} = 0\) when
\({\ell} \neq {\ell'}\). This maximization can be framed as
\begin{equation}
{{\bf L}_{\bf X}^{T}}{\bf L}_{\bf Y} = {\bf \Delta},
\end{equation} where \({\bf \Delta}\) is the diagonal matrix of singular
values, and so \({\bf \Delta}^{2} = {\bf \Lambda}\) which are
eigenvalues. Like with the GSVD, the GPLSSVD decomposes the relationship
between two data matrices into three matrices as \begin{equation}
[({\bf M}_{\bf X}^{\frac{1}{2}}{\bf X})^{T}({\bf M}_{\bf Y}^{\frac{1}{2}}{\bf Y})] = {\bf P}{\bf \Delta}{\bf Q}^{T},
\end{equation} where \({\bf \Delta}\) is the diagonal matrix of singular
values, and where \({\bf P}\) and \({\bf Q}\) are the left and right
\emph{generalized} singular vectors, respectively. Like the GSVD and
GEVD, the GPLSSVD finds orthogonal slices of
\(({\bf M}_{\bf X}^{\frac{1}{2}}{\bf X})^{T}({\bf M}_{\bf Y}^{\frac{1}{2}}{\bf Y})\)
with respect to the column constraints. The GPLSSVD maximizes
\({\bf \Delta} = {\bf P}^{T}{\bf W}_{\bf X}[({\bf M}_{\bf X}^{\frac{1}{2}}{\bf X})^{T}({\bf M}_{\bf Y}^{\frac{1}{2}}{\bf Y})]{\bf W}_{\bf Y}{\bf Q}\)
under the constraint of orthogonality where
\({\bf P}^{T}{\bf W}_{\bf X}{\bf P} = {\bf I} = {\bf Q}^{T}{\bf W}_{\bf Y}{\bf Q}\).
Typically, the GPLSSVD is performed with the SVD as \begin{equation}
\widetilde{\bf X}^{T}\widetilde{\bf Y} = {\bf U}{\bf \Delta}{\bf V},
\end{equation} where
\(\widetilde{\bf X} = {{\bf M}_{\bf X}^{\frac{1}{2}}}{\bf X}{{\bf W}^{\frac{1}{2}}_{\bf X}}\)
and
\(\widetilde{\bf Y} = {{\bf M}_{\bf Y}^{\frac{1}{2}}}{\bf Y}{{\bf W}^{\frac{1}{2}}_{\bf Y}}\),
and where \({\bf U}\) and \({\bf V}\) are the left and right singular
vectors, respectively, which are orthonormal such that
\({\bf U}^{T}{\bf U} = {\bf I} = {\bf V}^{T}{\bf V}\). The relationship
between the generalized and standard singular vectors are
\begin{equation}
\begin{aligned}
{\bf P} = {{\bf W}^{-\frac{1}{2}}_{\bf X}}{\bf U} \Longleftrightarrow {\bf U} = {{\bf W}^{\frac{1}{2}}_{\bf X}}{\bf P} \\
{\bf Q} = {{\bf W}^{-\frac{1}{2}}_{\bf Y}}{\bf V} \Longleftrightarrow {\bf V} = {{\bf W}^{\frac{1}{2}}_{\bf Y}}{\bf Q}.
\end{aligned}
\end{equation} When all constraint matrices are \({\bf I}\), the GPLSSVD
produces exactly the same results as the PLSSVD because
\(\widetilde{\bf X} = {\bf X}\) and \(\widetilde{\bf Y} = {\bf Y}\) and
thus \({\bf P} = {\bf U}\) and \({\bf Q} = {\bf V}\).

The latent variables are then expressed with respect to the constraints
and \emph{generalized} singular vectors as
\({\bf L}_{\bf X} = ({\bf M}_{\bf X}^{\frac{1}{2}}{\bf X}{\bf W}_{\bf X}{\bf P})\)
and
\({\bf L}_{\bf Y} = ({\bf M}_{\bf Y}^{\frac{1}{2}}{\bf Y}{\bf W}_{\bf Y}{\bf Q})\).
These latent variables maximize the weighted covariance (by way of the
constraints) subject to orthogonality where \begin{equation}
\begin{aligned}
{\bf L}_{\bf X}^{T}{\bf L}_{\bf Y} = \\
({\bf M}_{\bf X}^{\frac{1}{2}}{\bf X}{\bf W}_{\bf X}{\bf P})^{T}({\bf M}_{\bf Y}^{\frac{1}{2}}{\bf Y}{\bf W}_{\bf Y}{\bf Q}) =\\
(\widetilde{\bf X}{\bf U})^{T}(\widetilde{\bf Y}{\bf V}) =\\
{\bf U}^{T}\widetilde{\bf X}^{T}\widetilde{\bf Y}{\bf V} = {\bf \Delta}.
\end{aligned}
\end{equation}

We will see in the following section that the ``weighted covariance''
could be the correlation, which allows us to use the GPLSSVD to perform
various types of ``cross-decomposition'' techniques. Like with the GEVD
and GSVD, the GPLSSVD produces component or factor scores. The component
scores are defined as \begin{equation}
{\bf F}_{I} = {\bf W}_{\bf X}{\bf P}{\bf \Delta} \textrm{ and } {\bf F}_{J} = {\bf W}_{\bf Y}{\bf Q}{\bf \Delta},
\end{equation} for the columns of \({\bf X}\) and the columns of
\({\bf Y}\), respectively. The optimization in the GPLSSVD can be
reframed as the maximization of the component scores where
\({\bf F}_{I}^{T}{\bf W}_{\bf X}^{-1}{\bf F}_{I} = {\bf \Lambda} = {\bf F}_{J}^{T}{\bf W}_{\bf Y}^{-1}{\bf F}_{J}\)
where \({\bf \Lambda}\) are the eigenvalues, and this maximization is
still subject to
\({\bf P}^{T}{\bf W}_{\bf X}{\bf P} = {\bf I} = {\bf Q}^{T}{\bf W}_{\bf Y}{\bf Q}\).

\hypertarget{decomposition-tuples}{%
\subsection{Decomposition tuples}\label{decomposition-tuples}}

For simplicity, the GSVD is often referred to as a ``triplet'' or ``the
GSVD triplet'' \citep{husson_jan_2016, holmes_multivariate_2008}
comprised of (1) the data matrix, (2) the column constraints, and (3)
the row constraints. We can use the same concept to also define
``tuples'' for the GEVD and GPLSSVD. To note, the traditional way to
present the GSVD triplet is in the above order (data, column
constraints, row constraints). However, here I present a different order
for the elements in the tuples so that I can (1) better harmonize the
tuples across the three decompositions presented here, and (2) simplify
the tuples such that the order of the elements within the tuples
reflects the matrix multiplication steps. Furthermore, I present two
different tuples for each decomposition---a complete and a
partial---where the partial is a lower rank solution. The complete
decomposition tuples are:

\begin{itemize}
\item
  The complete GEVD 2-tuple:
  \(\mathrm{GEVD(}{\bf X}, {\bf W}_{\bf X}\mathrm{)}\)
\item
  The complete GSVD decomposition 3-tuple:
  \(\mathrm{GSVD(}{\bf M}_{\bf X}, {\bf X}, {\bf W}_{\bf X}\mathrm{)}\)
\item
  The complete GPLSSVD decomposition 6-tuple:
  \(\mathrm{GPLSSVD(}{\bf M}_{\bf X}, {\bf X}, {\bf W}_{\bf X}, {\bf M}_{\bf Y}, {\bf Y}, {\bf W}_{\bf Y}\mathrm{)}\).
\end{itemize}

Additionally, we can take the idea of tuples one step further and allow
for the these tuples to also define the desired \emph{returned rank} of
the results referred to as ``partial decompositions''. The partial
decompositions produce (return) only the first \(C\) components, and are
defined as:

\begin{itemize}
\item
  The partial GEVD decomposition 3-tuple:
  \(\mathrm{GEVD(}{\bf X}, {\bf W}_{\bf X}, C\mathrm{)}\)
\item
  The partial GSVD decomposition 4-tuple:
  \(\mathrm{GSVD(}{\bf M}_{\bf X}, {\bf X}, {\bf W}_{\bf X}, C\mathrm{)}\)
\item
  The partial GPLSSVD decomposition 7-tuple:
  \(\mathrm{GPLSSVD(}{\bf M}_{\bf X}, {\bf X}, {\bf W}_{\bf X}, {\bf M}_{\bf Y}, {\bf Y}, {\bf W}_{\bf Y}, C\mathrm{)}\).
\end{itemize}

Overall, these tuples provide short and convenient ways to express the
decompositions. And as we will see in later sections, these tuples
provide a simpler way to express specific techniques under the same
framework (e.g., PLS and CCA via GPLSSVD).

\hypertarget{restrictions-and-limitations}{%
\subsection{Restrictions and
limitations}\label{restrictions-and-limitations}}

In general, \pkg{GSVD} was designed around the most common analyses that
use the GSVD, GEVD, and GPLSSVD. These techniques are, typically,
multidimensional scaling (GEVD), principal components analysis (GEVD or
GSVD), correspondence analysis (GSVD), partial least squares (GSVD or
GPLSSVD), reduced rank regression (GSVD or GPLSSVD), canonical
correlation analysis (GSVD or GPLSSVD), and numerous variants and
extensions of all the aforementioned (and more).

One of the restrictions of these generalized decompositions is that any
constraint matrix must be positive semi-definite. That means, typically,
these matrices are square symmetric matrices with non-negative
eigenvalues. Often that means constraint matrices are, for examples,
covariance matrices or diagonal matrices. \pkg{GSVD} performs checks to
ensure adherence to positive semi-definiteness for constraint matrices.

Likewise, many techniques performed through the GEVD or EVD also assume
positive semi-definiteness. For example, PCA is the EVD of a correlation
or covariance matrix. Thus in \pkg{GSVD}, there are checks for positive
semi-definite matrices in the GEVD. Furthermore, all decompositions in
\pkg{GSVD} check for eigenvalues below a precision level. When found,
these very small eigenvalues are effectively zero and not returned by
any of the decompositions in \pkg{GSVD}. However, this can be changed
with an input parameter to allow for these vectors to be returned (I
discuss these parameters in the following section).

\hypertarget{other-and-related-decompositions.}{%
\subsection{Other and related
decompositions.}\label{other-and-related-decompositions.}}

To note, the GSVD discussed here is not the same as another technique
referred to as the ``generalized singular value decomposition''
\citep{van_loan_generalizing_1976}. The Van Loan generalized SVD has
more recently been referred to as the ``quotient SVD (QSVD)'' to help
distinguish it from the GSVD defined here
\citep{takane_relationships_2003}. Furthermore, there are numerous other
variants of the EVD and SVD beyond the GEVD and GSVD presented here.
\citet{takane_relationships_2003} provides a detailed explanation of
those variants as well as the relationships between the variants.

\hypertarget{other-packages-of-note}{%
\subsection{Other packages of note}\label{other-packages-of-note}}

There are multiple packages that implement methods based on the GSVD or
GEVD, where some of do in fact make use of a GSVD or GSVD-like call.
Generally, though, the GSVD calls in these packages are meant more for
internal (to the package) use instead of a more external facing tool
like \pkg{GSVD}. A core (but by no means comprehensive) list of those
packages follows. There are at least four packages designed to provide
interfaces to specific GSVD techniques, and each of those packages has
an internal function meant for GSVD-like decomposition.

\begin{itemize}
\item
  \pkg{ExPosition} which includes the function \texttt{genPDQ}
  \citep{beaton_exposition_2014}. As the author of \pkg{ExPosition}, I
  designed \texttt{genPDQ} under the most common usages which are
  generally diagonal matrices (or simply just vectors). I regret that
  design choice, which is one of the primary motivations why I developed
  \pkg{GSVD}.
\item
  \pkg{FactoMineR} \citep{le_factominer_2008} includes the function
  \texttt{svd.triplet}, which also makes use of vectors instead of
  matrices because the most common (conceptual) uses of the GSVD is that
  the row and/or column constraints are diagonal matrices.
\item
  \pkg{ade4} \citep{dray_ade4_2007} which includes a function called
  \texttt{as.dudi} that is the core decomposition step. It, too, makes
  use of vectors.
\item
  \pkg{amap} \citep{lucas_amap_2019} which includes the function
  \texttt{acp} which, akin to the previous 3 packages listed, also only
  makes use of vectors for row and column constraints.
\end{itemize}

However, there are other packages that also more generally provide
methods based on GSVD, GEVD, or GPLSSVD approaches however, they do not
necessarily include a GSVD-like decomposition. Instead they make more
direct use of the SVD or eigendecompositions, or alternative methods
(e.g., alternating least squares). Those packages include but are not
limited to:

\begin{itemize}
\item
  \pkg{MASS} \citep{venables_modern_2002} includes
  \texttt{MASS::corresp} which is an implementation of CA,
\item
  \pkg{ca} which includes a number of CA-based methods
  \citep{nenadic_correspondence_2007},
\item
  \pkg{CAvariants} is another package that includes standard CA and
  other variations not seen in other packages
  \citep{lombardo_variants_2016},
\item
  \pkg{homals}, and \pkg{anacor}, are packages each that address a
  number of methods in the CA family
  \citep{leeuw_gifi_2009, de_leeuw_simple_2009},
\item
  \pkg{candisc} focuses on canonical discriminant and correlation
  analyses \citep{friendly_candisc_2020},
\item
  \pkg{vegan} that includes a very large set of ordination methods with
  a particular emphasis on cross-decomposition or canonical methods
  \citep{oksanen_vegan_2019}, and
\item
  \pkg{rgcca} also presents a more unified approach to CCA and PLS under
  a more unified framework that also includes various approaches to
  regularization
  \citep{tenenhaus_rgcca_2017, tenenhaus_regularized_2014}.
\item
  \pkg{ics} is a package for invariant coordinate selection, which can
  obtain unmixing matrices (i.e., as in independent components analysis)
  \citep{nordhausen_tools_2008, tyler_invariant_2009}.
\end{itemize}

\hypertarget{package-description-core-functions-and-features}{%
\section{Package description: core functions and
features}\label{package-description-core-functions-and-features}}

The \pkg{GSVD} package has three primary ``workhorse'' functions:

\begin{itemize}
\item
  \code{geigen(X, W, k = 0, tol = sqrt(.Machine\$double.eps), symmetric)},
\item
  \code{gsvd(X, LW, RW, k = 0, tol = .Machine\$double.eps)}, and
\item
  \code{gplssvd(X, Y, XLW, YLW, XRW, YRW, k = 0, tol = .Machine\$double.eps)}
\end{itemize}

In \code{geigen()} or \code{gsvd()} each there is one data matrix
\code{X}, whereas \code{gplssvd()} has two data matrices \code{X} and
\code{Y}. In \code{geigen()} there is a single constraint matrix
\code{W}. In \code{gsvd()} there are two constraint matrices, \code{LW}
or ``left weights'' for the rows of \code{X} and \code{RW} or ``right
weights'' for the columns of \code{X}. The ``left'' and ``right''
references are used because of the association between these weights and
the left and right generalized singular vectors. In \code{gplssvd()}
there are two constraint matrices per data matrix (so four total
constraint matrices): \code{XLW} and \code{XRW} for \code{X}'s ``left''
and ``right'' weights, and \code{YLW} and \code{YRW} for \code{Y}'s
``left'' and ``right'' weights. The \code{geigen()} includes the
argument \code{symmetric} to indicate if \code{X} is a symmetric matrix;
when missing \code{X} is tested via \code{isSymmetric()}. The
\code{symmetric} argument is eventually passed through to, and is the
same as, \code{symmetric} in \code{base::eigen()}. All three functions
include \code{k} which indicates how many components to return. Finally,
all three functions include a tolerance argument \code{tol}, which is
passed through to \code{tolerance_svd()} or \code{tolerance_eigen()}.
These functions are the same as \code{base::svd()} and
\code{base::eigen()}, respectively, with the added tolerance feature. In
both cases, the \code{tol} argument is used to check for any eigenvalues
or singular values below the tolerance threshold. Any eigen- or singular
values below that threshold are discarded, as they are effectively zero.
These values occur when data are collinear, which is common in high
dimensional cases or in techniques such as Multiple Correspondence
Analysis. However, the \code{tol} argument can be effectively turned off
with the use of \code{NA}, \code{NULL}, \code{Inf}, \code{-Inf},
\code{NaN}, or any value \(< 0\). In this case, both
\code{tolerance_svd()} and \code{tolerance_eigen()} simply call
\code{base::svd()} and \code{base::eigen()} with no changes. When using
the \code{tol} argument, eigen- and singular values are also checked to
ensure that they are real and positive values. If they are not, then
\code{geigen()}, \code{gsvd()}, and \code{gplssvd()} stop. The
motivation behind this behavior is because the \code{geigen()},
\code{gsvd()}, and \code{gplssvd()} functions are meant to perform
routine multivariate analyses---such as MDS, PCA, CA, CCA, or PLS---that
require data and/or constraint matrices assumed to be positive
semi-definite.

Data matrices are the minimally required objects for \code{geigen()},
\code{gsvd()}, and \code{gplssvd()}. All other arguments (input) either
have suitable defaults or are allowed to be missing. For example, when
any of the constraints (``weights'') are missing, then the constraints
are mathematically equivalent to identity matrices (i.e., \({\bf I}\))
which contain \(1\)s on the diagonal with \(0\)s off-diagonal. Table
\ref{tab:arguments} shows a mapping between our (more formal) notation
above and our more intuitively named arguments for the functions. The
rows of Table \ref{tab:arguments} are the three primary
functions---\code{geigen()}, \code{gsvd()}, and \code{gplssvd()}---where
the columns are the elements used in the formal notation (and also used
in the tuple notation).

\begin{table}
\begin{center}
\begin{tabular}{l*{6}{c}r}
   & ${\bf X}$ & ${\bf Y}$ & ${{\bf M}_{\bf X}}$ & ${{\bf W}_{\bf X}}$ & ${{\bf M}_{\bf Y}}$ & ${{\bf W}_{\bf Y}}$ & $C$ \\
\hline
\code{geigen()}  & \code{X} & - & - & \code{W} & - & - & \code{k}  \\
\code{gsvd()}  & \code{X} & - & \code{LW} & \code{RW} & - & - &  \code{k}  \\
\code{gplssvd()}  & \code{X} & \code{Y} & \code{XRW} & \code{XLW} & \code{YRW} & \code{YLW} & \code{k}  \\
\end{tabular}
\caption{\label{tab:arguments}Mapping between arguments (input) to functions (rows) and notation for the analysis tuples (columns).}
\end{center}
\end{table}

Additionally, there are some ``helper'' and convenience functions used
internally to the \code{geigen()}, \code{gsvd()}, and \code{gplssvd()}
functions that are made available for use. These include
\code{sqrt_psd_matrix()} and \code{invsqrt_psd_matrix()} which compute
the square root (\code{sqrt}) and inverse square root (\code{invsqrt})
of positive semi-definite (\code{psd}) matrices (\code{matrix}),
respectively. The \pkg{GSVD} package also includes helpful functions for
testing matrices: \code{is_diagaonal_matrix()} and
\code{is_empty_matrix()}. Both of these tests help minimize the memory
and computational footprints for, or check validity of, the constraints
matrices.

Finally, the three core functions in \pkg{GSVD}---\code{geigen()},
\code{gsvd()}, and \code{gplssvd()}---each have their own class objects
but provide overlapping and identical outputs. The class object is
hierarchical from a list, to a package, to the specific function:
\code{c("geigen","GSVD","list")}, \code{c("gsvd","GSVD","list")}, and
\code{c("gplssvd","GSVD","list")} for \code{geigen()}, \code{gsvd()},
and \code{gplssvd()} respectively. Table \ref{tab:values} list the
possible outputs across \code{geigen()}, \code{gsvd()}, and
\code{gplssvd()}. The first column of Table \ref{tab:values} explains
the returned value, where the second column provides a mapping back to
the notation used here. The last three columns indicate---with an
`X'---which of the returned values are available from the \code{geigen},
\code{gsvd}, or \code{gplssvd} functions.

\begin{table}
\begin{center}
\begin{tabular}{l*{4}{c}r}
   & What it is & Notation & \code{geigen} & \code{gsvd} & \code{gplssvd} \\
\hline
\code{d} & \code{k} singular values & ${\bf \Delta}$ & \checkmark & \checkmark & \checkmark\\
\code{d_full} & all singular values & ${\bf \Delta}$ & \checkmark & \checkmark & \checkmark \\
\code{l} & \code{k} eigenvalues & ${\bf \Lambda}$ & \checkmark & \checkmark & \checkmark \\
\code{l_full} & all eigenvalues & ${\bf \Lambda}$ & \checkmark & \checkmark & \checkmark \\
\code{u} & \code{k} Left singular/eigen vectors & ${\bf U}$ &  & \checkmark & \checkmark \\
\code{v} & \code{k} Right singular/eigen vectors & ${\bf V}$ & \checkmark & \checkmark & \checkmark \\
\code{p} & \code{k} Left generalized singular/eigen vectors & ${\bf P}$ & & \checkmark & \checkmark \\
\code{q} & \code{k} Right generalized singular/eigen vectors & ${\bf Q}$ & \checkmark & \checkmark & \checkmark \\
\code{fi} & \code{k} Left component scores & ${\bf F}_{I}$ & & \checkmark & \checkmark \\
\code{fj} & \code{k} Right component scores & ${\bf F}_{J}$ & \checkmark & \checkmark & \checkmark \\
\code{lx} & \code{k} Latent variable scores for \code{X} & ${\bf L}_{\bf X}$ & & & \checkmark \\
\code{ly} & \code{k} Latent variable scores for \code{Y} & ${\bf L}_{\bf Y}$ & & & \checkmark \\
\end{tabular}
\caption{\label{tab:values}Mapping of values (output from functions; rows) to their conceptual meanings, notation used here, and which \pkg{GSVD} functions have these values.}
\end{center}
\end{table}

Different fields and traditions use different nomenclature, different
descriptions, or different ways of framing optimizations. But
conceptually and mathematically, numerous multivariate techniques are
much more related than they appear, especially when solved with the EVD
and SVD. The \pkg{GSVD} package provides a single framework to unify
common multivariate analyses by way of three generalized decompositions.
The \code{arguments} (function inputs) and \code{values} (function
outputs) help reinforce the mathematical and coneptual equivalence and
relationships between techniques, and now via the \pkg{GSVD} package, we
see this unification programatically and through analyses. Therefore the
common names---across the core functions---for the \code{values} in
Table \ref{tab:values} was an intentional design choice.

Finally, this package is ``lightweight'' in that it is written in base
R, with no dependencies, and has a minimal set of functions to achieve
the goals of \pkg{GSVD}. Furthermore, a number of strategies were
employed in order to minimize both memory and computational footprints.
For example, when constraints matrices are available, they are checked
for certain conditions. Specifically, if a matrix is a diagonal matrix
is it transformed into a vector, which decreases memory consumption and
speeds up some computations (e.g., multiplication). If that same vector
is all \(1\)s, then the matrix was an identity matrix, and is then
ignored in all computation.

\hypertarget{examples-of-multivariate-analyses}{%
\section{Examples of multivariate
analyses}\label{examples-of-multivariate-analyses}}

In this section, I present many of the most commonly used multivariate
analyses, and how they can be performed through the \pkg{GSVD} package,
as well as how these methods can be framed in various ways. Here, I
focus primarily on what are likely the most common: principal components
analysis, multidimensional scaling, correspondence analysis (and some of
its variations), partial least squares, reduced rank regression
\citep[a.k.a. redundnacy analysis or multivariable multivariate
regression;][]{van1977redundancy, de2012least}, and canonical
correlation analysis. As I introduce these methods, I also introduce how
to use various functions (and their parameters) in \pkg{GSVD}.

There are other very common multivariate techniques, such as log and
power CA methods
\citep{greenacre_correspondence_2010, greenacre2009power}, and various
discriminant analyses. I forgo the descriptions of these latter cases
because they tend to be specific or special cases of techniques I do
highlight. For examples: log-linear CA requires additional
transformations and then performs CA, and various discriminant analyses
reduce to special cases of PLS, RRR, or CCA.

The \pkg{GSVD} package contains several toy or illustrative data sets
that work as small examples of various techniques. There is also a
larger and more realistic data set in the package that I use in the
following examples. That data set is a synthetic data set modeled after
data from the Ontario Neurodegenerative Disease Research Initiative
(ONDRI; \url{https://ondri.ca/}). The synthetic data were generated from
real data, but were ``synthesized'' with the \pkg{synthpop} package
\citep{synthpop}. This synthetic data
set---\code{synthetic_ONDRI}---contains 138 rows (observations) and 17
columns (variables). See \url{https://github.com/ondri-nibs} for more
details. The \code{synthetic_ONDRI} data set is particularly useful
because it contains a variety of data types, e.g., some quantitative
such as cognitive or behavioral scores that are continuous, and brain
volumes that are strictly positive integers, as well as categorical and
ordinal variables (typically demographic or clinical measures). For each
of the following illustrations of techniques, I use particular subsets
of the \code{synthetic_ONDRI} data most relevant or appropriate for
those techniques (e.g., continuous measures for PCA, distances for MDS,
cross-tabulations of categories for CA).

\hypertarget{principal-components-analysis}{%
\subsection{Principal components
analysis}\label{principal-components-analysis}}

Generally, there are two ways to approach PCA: with a covariance matrix
or with a correlation matrix. First, I show both of these PCA approaches
on a subset of continuous measures from the \code{synthetic_ONDRI}
dataset. Then I focus on correlation PCA, but with an emphasis on (some
of) the variety of ways we can perform correlation PCA with generalized
decompositions. PCA is illustrated with a subset of continuous measures
from cognitive tasks.

We can perform a covariance PCA and a correlation PCA with the
generalized eigendecomposition as:

\begin{CodeChunk}
\begin{CodeInput}
R> continuous_data <- synthetic_ONDRI[,c("TMT_A_sec", "TMT_B_sec",
+                                       "Stroop_color_sec", "Stroop_word_sec", 
+                                       "Stroop_inhibit_sec", "Stroop_switch_sec")]
R> 
R> cov_pca_geigen <- geigen( cov(continuous_data) )
R> cor_pca_geigen <- geigen( cor(continuous_data) )
\end{CodeInput}
\end{CodeChunk}

In these cases, the use here is no different---from a user
perspective---of how PCA would be performed with the plain \code{eigen}.
For now, the major advantage of the \code{geigen} approach is that the
output (values) also include component scores and other measures common
to these decompositions, such as singular values, as seen in the output.
The following code chunk shows the results of the \code{print} method
for \code{geigen}

\begin{CodeChunk}
\begin{CodeInput}
R> cov_pca_geigen
\end{CodeInput}
\begin{CodeOutput}
**GSVD package object of class type 'geigen'.**

geigen() was performed on a marix with 6 columns/rows
Number of total components = 6.
Number of retained components = 6.

The 'geigen' object contains:                                            
 $d_full Full set of singular values        
 $l_full Full set of eigen values           
 $d      Retained set of singular values (k)
 $l      Retained set of eigen values (k)   
 $v      Eigen/singular vectors             
 $q      Generalized eigen/singular vectors 
 $fj     Component scores                   
\end{CodeOutput}
\begin{CodeInput}
R> cor_pca_geigen
\end{CodeInput}
\begin{CodeOutput}
**GSVD package object of class type 'geigen'.**

geigen() was performed on a marix with 6 columns/rows
Number of total components = 6.
Number of retained components = 6.

The 'geigen' object contains:                                            
 $d_full Full set of singular values        
 $l_full Full set of eigen values           
 $d      Retained set of singular values (k)
 $l      Retained set of eigen values (k)   
 $v      Eigen/singular vectors             
 $q      Generalized eigen/singular vectors 
 $fj     Component scores                   
\end{CodeOutput}
\end{CodeChunk}

A more comprehensive approach to PCA is the analysis of a rectangular
table---i.e., the observations by the measures---as opposed to the
square symmetric matrix of relationships between the variables. The more
comprehensive approach to PCA can be performed with \code{gsvd}, as
shown below, with its \code{print} method to show the output from
\code{gsvd}. For this example of correlation PCA, we introduce a set of
constraints for the rows. For standard PCA, the row constraints are just
\(\frac{1}{I}\) where \(I\) is the number of rows in the data matrix.

\begin{CodeChunk}
\begin{CodeInput}
R> scaled_data <- scale(continuous_data, center = T, scale = T)
R> degrees_of_freedom <- nrow(continuous_data)-1
R> row_constraints <- rep(1/(degrees_of_freedom), nrow(scaled_data))
R> 
R> cor_pca_gsvd <- gsvd( scaled_data, LW = row_constraints )
R> 
R> cor_pca_gsvd
\end{CodeInput}
\begin{CodeOutput}
**GSVD package object of class type 'gsvd'.**

gsvd() was performed on a matrix with 138 rows and 6 columns
Number of components = 6.
Number of retained components = 6.

The 'gsvd' object contains:                                                                
 $d_full Full set of singular values                            
 $l_full Full set of eigen values                               
 $d      Retained set of singular values (k)                    
 $l      Retained set of eigen values (k)                       
 $u      Left singular vectors (for rows of DAT)                
 $v      Right singular vectors (for columns of DAT)            
 $p      Left generalized singular vectors (for rows of DAT)    
 $q      Right generalized singular vectors (for columns of DAT)
 $fi     Left component scores (for rows of DAT)                
 $fj     Right component scores (for columns of DAT)            
\end{CodeOutput}
\end{CodeChunk}

In the next code chunk, we can see that the results for the common
objects---the singular and eigenvalues, the generalized and plain
singular vectors, and the component scores---are all identical between
\code{geigen} and \code{gsvd}.

\begin{CodeChunk}
\begin{CodeInput}
R> all(
+ all.equal(cor_pca_geigen$l_full, cor_pca_gsvd$l_full),
+ all.equal(cor_pca_geigen$d_full, cor_pca_gsvd$d_full),
+ all.equal(cor_pca_geigen$v, cor_pca_gsvd$v),
+ all.equal(cor_pca_geigen$q, cor_pca_gsvd$q),
+ all.equal(cor_pca_geigen$fj, cor_pca_gsvd$fj)
+ )
\end{CodeInput}
\begin{CodeOutput}
[1] TRUE
\end{CodeOutput}
\end{CodeChunk}

The use of row constraints helps illustrate the GSVD triplet. The
previous correlation PCA example is
\(\mathrm{GSVD(}\frac{1}{I-1}{\bf I}, {\bf X}, {\bf I}\mathrm{)}\) where
\(\frac{1}{I}{\bf I}\) is a diagaonal matrix of \(\frac{1}{I-1}\) with
\(I\) as the number of rows of \({\bf X}\) and \({\bf I}\) is the
identity matrix. In this particular case, the row constraints are all
equal and thus a simple scaling factor. Alternatively, we could just
divide by that scalar as
\code{gsvd( scaled_data / sqrt(degrees_of_freedom))} which is equivalent
to \code{geigen( cor(continuous_data) )}.

The previous PCA
triplet---\(\mathrm{GSVD(}\frac{1}{I}{\bf I}, {\bf X}, {\bf I}\mathrm{)}\)---helps
us transition to an alternative view of correlation PCA and an expanded
use of the GSVD triplet. Correlation PCA can be reframed as covariance
PCA with additional constraints imposed on the columns. Let's first
compute PCA with row and column constraints via \code{gsvd}, then go
into more detail on the GSVD triplet and its computation.

\begin{CodeChunk}
\begin{CodeInput}
R> centered_data <- scale(continuous_data, scale = F)
R> degrees_of_freedom <- nrow(centered_data)-1
R> row_constraints <- rep(1/(degrees_of_freedom), nrow(centered_data))
R> col_constraints <-  degrees_of_freedom / colSums( (centered_data)^2 )
R> 
R> cor_pca_gsvd_triplet <- gsvd(centered_data, 
+                              LW = row_constraints,  
+                              RW = col_constraints)
\end{CodeInput}
\end{CodeChunk}

First, let's test the equality between the previous correlation PCA and
this alternative approach.

\begin{CodeChunk}
\begin{CodeInput}
R> mapply(all.equal, cor_pca_gsvd, cor_pca_gsvd_triplet)
\end{CodeInput}
\begin{CodeOutput}
                                    d                                     u 
                               "TRUE"                                "TRUE" 
                                    v                                d_full 
                               "TRUE"                                "TRUE" 
                               l_full                                     l 
                               "TRUE"                                "TRUE" 
                                    p                                    fi 
                               "TRUE"                                "TRUE" 
                                    q                                    fj 
 "Mean relative difference: 23.65976" "Mean relative difference: 0.9420834" 
\end{CodeOutput}
\end{CodeChunk}

We can see that almost everything is identical, with two exceptions:
\code{q} and \code{fj}. That's because the decomposed matrix is
identical across the two, but their constraints are not. The
commonalities and the differences stem from how the problem was framed.
Let's unpack what we did for the three versions of PCA shown. Let's
assume that \({\bf X}\) is a matrix that has been column-wise centered.
The covariance matrix for \({\bf X}\) is
\({\bf \Sigma} = {\bf X}^{T}(\frac{1}{I-1}{\bf I}){\bf X}\), where
\({\bf d} = \mathrm{diag\{}{\bf \Sigma}\mathrm{\}}\), which are the
variances per variable, and
\({\bf D} = \mathrm{diag\{}{\bf d}\mathrm{\}}\) which is a diagonal
matrix of the variances. Each of the three correlation PCA approaches
can be framed as the following generalized decompositions:

\begin{itemize}
\item
  \(\mathrm{GEVD(}{\bf D}^{-\frac{1}{2}}{\bf S}{\bf D}^{-\frac{1}{2}}, {\bf I}\mathrm{)}\)
\item
  \(\mathrm{GSVD(}\frac{1}{I-1}{\bf I}, {\bf X}{\bf D}^{-\frac{1}{2}}, {\bf I}\mathrm{)}\)
\item
  \(\mathrm{GSVD(}\frac{1}{I-1}{\bf I}, {\bf X}, {\bf D}^{-1} \mathrm{)}\)
\end{itemize}

The second way of framing
PCA---\(\mathrm{GSVD(}\frac{1}{I-1}{\bf I}, {\bf X}{\bf D}^{-\frac{1}{2}}, {\bf I}\mathrm{)}\)---is
the canonical ``French way'' of presenting PCA via the GSVD: a centered
data matrix (\({\bf X}\)) normalized/scaled by the square root of the
variances (\({\bf D}^{-\frac{1}{2}}\)), with equal weights of
\({\frac{1}{I}}\) for the rows and an identity matrix for the columns.

Though not always practical for simple PCA computation, the use of
constraints in a GEVD doublet and GSVD triplet shows that we can frame
(or re-frame) various techniques as decompositions of data with respect
to constraints. In the case of PCA, the the constraint-based
decompositions highlight that we could perform PCA with any (suitable)
constraints applied to the rows and columns. So for example, we may have
\emph{a priori} knowledge about the observations (rows) and elect to use
different weights for each observation (we'll actually see this in the
MDS examples). We could also use some known or assumed population
variance as column constraints instead of the sample variance, or use
column constraints as a set of weights to impose on the variables in the
data.

\hypertarget{metric-multidimensional-scaling}{%
\subsection{(Metric) Multidimensional
scaling}\label{metric-multidimensional-scaling}}

Metric multidimensional scaling (MDS) is a technique akin to PCA, but
specifically for the factorization of distance matrix
\citep{torgerson1952multidimensional, borg2005modern}. MDS, like PCA, is
also an eigen-technique. Like the PCA examples, I show several ways to
perform MDS through the generalized approaches; specifically all through
the GEVD. But for this particular example we will (eventually) make use
of some known or \emph{a priori} information as the constraints. First
we show how to perform MDS as a plain EVD problem and then with
constraints as a GEVD problem. In these MDS illustrations, I use a
conceptually simpler albeit computationally more expensive approach. For
example, I use centering matrices and matrix algebra (in \proglang{R})
to show the steps. Much more efficient methods exist. In these examples
we use the same measures as in the PCA example.

\begin{CodeChunk}
\begin{CodeInput}
R> data_for_distances <- synthetic_ONDRI[,
+                                       c("TMT_A_sec", "TMT_B_sec", 
+                                         "Stroop_color_sec", "Stroop_word_sec",
+                                         "Stroop_inhibit_sec","Stroop_switch_sec")]
R> scaled_data <- scale(data_for_distances, center = T, scale = T)
R> 
R> distance_matrix <- as.matrix( dist( scaled_data ) )
R> 
R> row_weights <- rep(1/nrow(distance_matrix), nrow(distance_matrix))
R> 
R> centering_matrix <- diag(nrow(distance_matrix)) - 
+   ( rep(1,nrow(distance_matrix)) %o% row_weights )
R> 
R> matrix_to_decompose <- centering_matrix %*% 
+   (-(distance_matrix^2)/2) %*% 
+   t(centering_matrix)
R> 
R> mds_geigen <- geigen(matrix_to_decompose)
\end{CodeInput}
\end{CodeChunk}

The results from \code{geigen(matrix_to_decompose)} produce a variety of
outputs that align with the concept of eigenvectors, generalized
eigenvectors, and component scores. But, more specifically, the results
of \code{base::cmdscale(distance_matrix)} are identical to
\code{mds_geigen$fj[,1:2]}; that is, MDS scores as viewed through
\code{geigen} are component scores.

However, the generalized approach allows us to include constraints. In
the following example, we can now include a weighting factor per
observation as a constraint to impose on that observation. Here we use
the inverse of age, so that we effectively downweight older individuals
and upweight younger individuals.

\begin{CodeChunk}
\begin{CodeInput}
R> row_weights <- 1/synthetic_ONDRI$AGE
R> 
R> centering_matrix <- 
+   diag(nrow(distance_matrix)) - (
+   rep(1,nrow(distance_matrix)) %o% ( row_weights/sum(row_weights) )
+ )
R> 
R> 
R> matrix_to_decompose <- 
+   -(centering_matrix %*% (distance_matrix^2) %*% t(centering_matrix))/2
R> 
R> mds_weighted_geigen <- geigen( matrix_to_decompose , W = row_weights, tol = NA )
\end{CodeInput}
\end{CodeChunk}

In \code{mds_weighted_geigen} we require the use of \code{tol=NA}. This
is because \code{matrix_to_decompose} is not positive semi-definite.
Recall that one of the key principles of the \pkg{GSVD} package is that
we require positive semi-definite matrices for the constraints, and that
the design of \code{geigen} also---by default---assume positive
semi-definite matrices. This is because most multivariate
analyses---from the eigendecomposition perspective---require correlation
or covariance matrices which are by definition positive semi-definite.
If we were to run \code{geigen(matrix_to_decompose, diag(row_weights))}
we would get an error. In fact, we are unable to set an appropriate
tolerance parameter in this case because the last few eigenvalues have
large \emph{negative} values. But the use of \code{tol=NA} allows for a
direct computation of the eigendecomposition, without dropping any of
the dimensions (e.g., those below tolerance). For such an analysis, it
is typical to discard dimensions with such eigenvalues (which is why it
is a default in the package). However, some standard analyses do violate
those assumptions. For examples: the weighted MDS here, principal axis
factoring, and other approaches to factor analysis where, for example,
the diagonal of a correlation matrix is loaded with the communalities.
So the \pkg{GSVD} package accommodates these approaches by effectively
ignoring the tolerance requirement for eigenvalues.

The weighted version of MDS produces the same kind of results as
\code{vegan::wcmdscale}, but there exist differences the computation of
scores. From the perspective of \pkg{GSVD} the scores are the
generalized singular vectors, scaled by the singular values per
component, and with respect to a metric (i.e., the constraints).
Thus---as we've seen in the previous section---the component scores are
\({\bf W}_{\bf X}{\bf Q}{\bf \Delta}\). Many SVD and eigen-based
analyses have numerous ways of presenting and visualizing results. In
fact, the \pkg{vegan} package presents a slightly different version of
the weighted MDS scores compared to \pkg{GSVD}. From the \pkg{GSVD}
perspective, \pkg{vegan} produces scores computed from the generalized
singular vectors scaled by their respective singular values, or
\({\bf Q}{\bf \Delta}\). We can see this in Figure \ref{fig:mds}, which
shows the first two components from MDS and the weighted MDS, with
component 1 on the horizontal axis and component 2 on the vertical axis.
Figures \ref{fig:mds}a, c, and e each show the standard MDS solution
where Figures \ref{fig:mds}b, d, and f show the weighted (constrained)
MDS solution. Figures \ref{fig:mds}a and b show the eigen/singular
vectors scaled by the singular values (\({\bf V}{\bf \Delta}\)), Figures
\ref{fig:mds}c and d show the weighted generalized eigen/singular
vectors scaled by the singular values
(\({\bf W}_{\bf X}{\bf Q}{\bf \Delta}\)), and Figures \ref{fig:mds}c and
d show the generalized eigen/singular vectors scaled by the singular
values (\({\bf Q}{\bf \Delta}\)). Note that, in general, component
scores (e.g., Figures \ref{fig:mds}e and f) should show axes proportion
to their eigen or singular values \citep{nguyen_ten_2019}. This is true
for virtually all analyses shown in this paper. However, for simplicity
and to help with comparisons, here I do not scale the axes. We can see
in Figure \ref{fig:mds} that the standard form of MDS produces sets of
scores that show no differences in their pattern, but that the weighted
version does across the various scores. Furthermore, we can also see
that each weighted version set provides a slightly different perspective
compared to the standard MDS.

\begin{CodeChunk}
\begin{figure}

{\centering \includegraphics{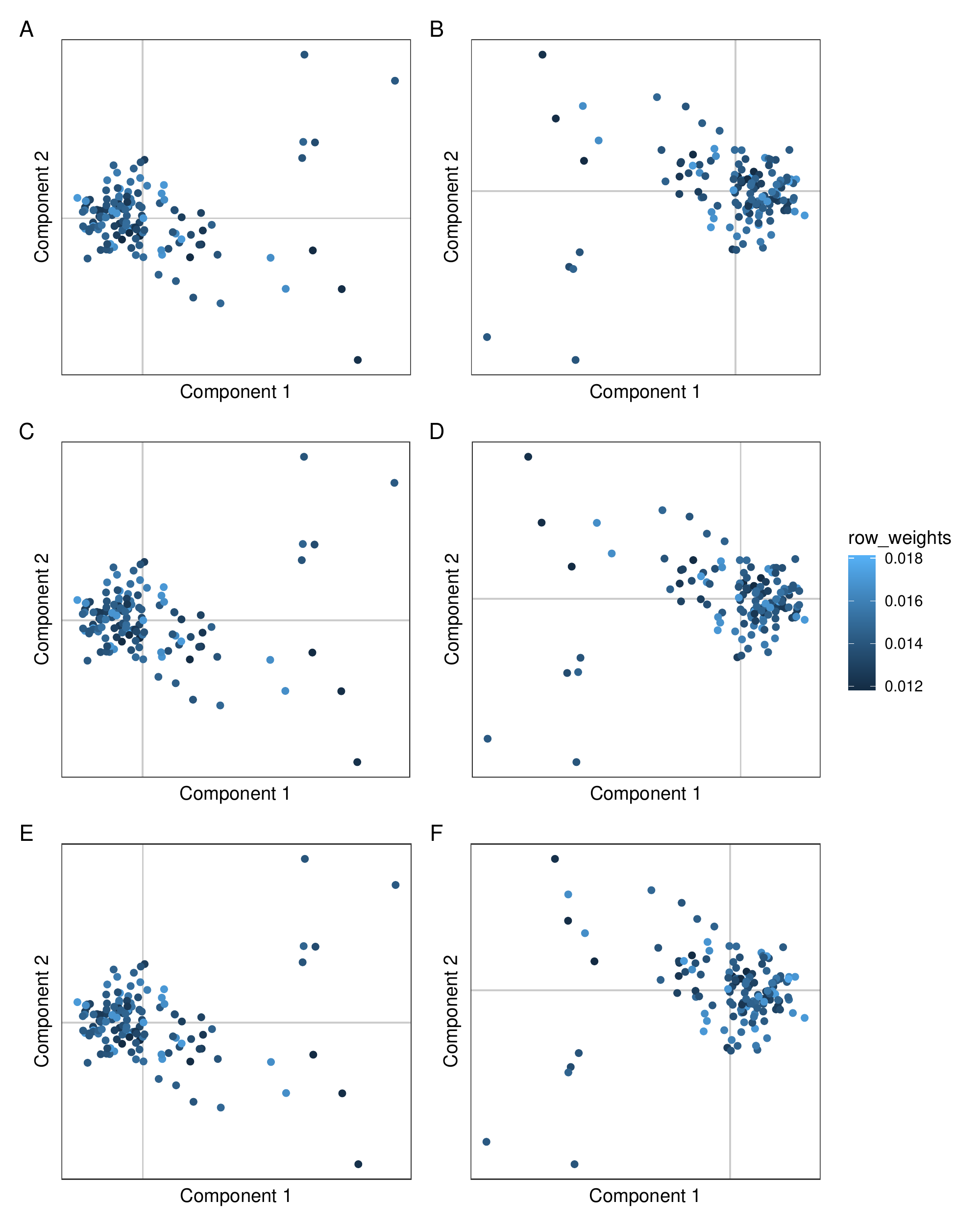} 

}

\caption{\label{fig:mds} Metric multidimensional scaling (MDS). Panels A, C, and E show standard MDS where panels B, D, and F show weighted MDS. Panels A and B show the singular vectors, panels C and D show the generalized singular vectors, and panels E and F show the component scores. Note how the configuration of the observations change in the weighted MDS through the different sets of scores (panels B, D, and F).}\label{fig:mds_vis}
\end{figure}
\end{CodeChunk}

\hypertarget{correspondence-analysis}{%
\subsection{Correspondence analysis}\label{correspondence-analysis}}

Correspondence analysis (CA) is one of---if not
\emph{the}---prototypical GSVD triplet method. CA is like PCA, but was
originally designed for two-way contingency tables, and was then
expanded into multiple correspondence analysis (MCA) for N-way
contingency tables
\citep{greenacre_theory_1984, lebart_multivariate_1984, escofier-cordier_analyse_1965, greenacre_correspondence_2010, greenacre2006multiple}.
MCA is more like PCA in that the rows are typically observations and the
columns are measures, where the data are transformed into ``complete
disjunctive coding'' (a.k.a. nominal coding, dummy coding, one-hot
encoding, and a variety of other names). CA methods can be thought of as
a ``\({\chi^2}\) PCA''. Prior to decomposition, the data matrix for CA
methods are preprocessed in a way to make the data table analogous to
that of a \({\chi^2}\) table, where we decompose the (weighted)
deviations under the assumption of independence.

The next section goes through three forms of CA. Each form helps
highlight the utility and flexibility of the GSVD and make use of
various forms of weights. Each illustration is one of the many
well-established derivatives of CA. The first example illustrates
standard CA (on a two-way contingency table). The second example
illustrates MCA on multiple categorical variables, which is then
followed by---and compared with---the third example, a
\emph{regularized} MCA of the same data.

Let's begin with standard CA---which is applied to a two-way contingency
table. Here we'll use genotypes from two genes: ApoE and MAPT. Both are
risk factors in neurodegenerative disorders.

\begin{CodeChunk}
\begin{CodeInput}
R> observed_matrix <- mapt_by_apoe_table / sum(mapt_by_apoe_table)
R> row_probabilities <- rowSums(observed_matrix)
R> col_probabilities <- colSums(observed_matrix)
R> expected_matrix <- row_probabilities %o% col_probabilities
R> deviations_matrix <- observed_matrix - expected_matrix
R> 
R> ca_gsvd <- gsvd( deviations_matrix, 
+                  LW = 1/row_probabilities, 
+                  RW = 1/col_probabilities )
\end{CodeInput}
\end{CodeChunk}

In CA, the primarily and almost exclusively visualized items are the
component scores: \code{fi} and \code{fj}. Recall that the components
scores are the generalized singular vectors, scaled by the singular
values, under the metric defined by the constraints, or
\({\bf M}_{\bf X}{\bf P}{\bf \Delta}\) and
\({\bf W}_{\bf X}{\bf P}{\bf \Delta}\). The reason behind the scores as
visuals in CA is that the scores reflect the \({\bf \chi}^2\) distance.
The CA component scores are shown in Figure \ref{fig:ca}.

\begin{CodeChunk}
\begin{figure}

{\centering \includegraphics{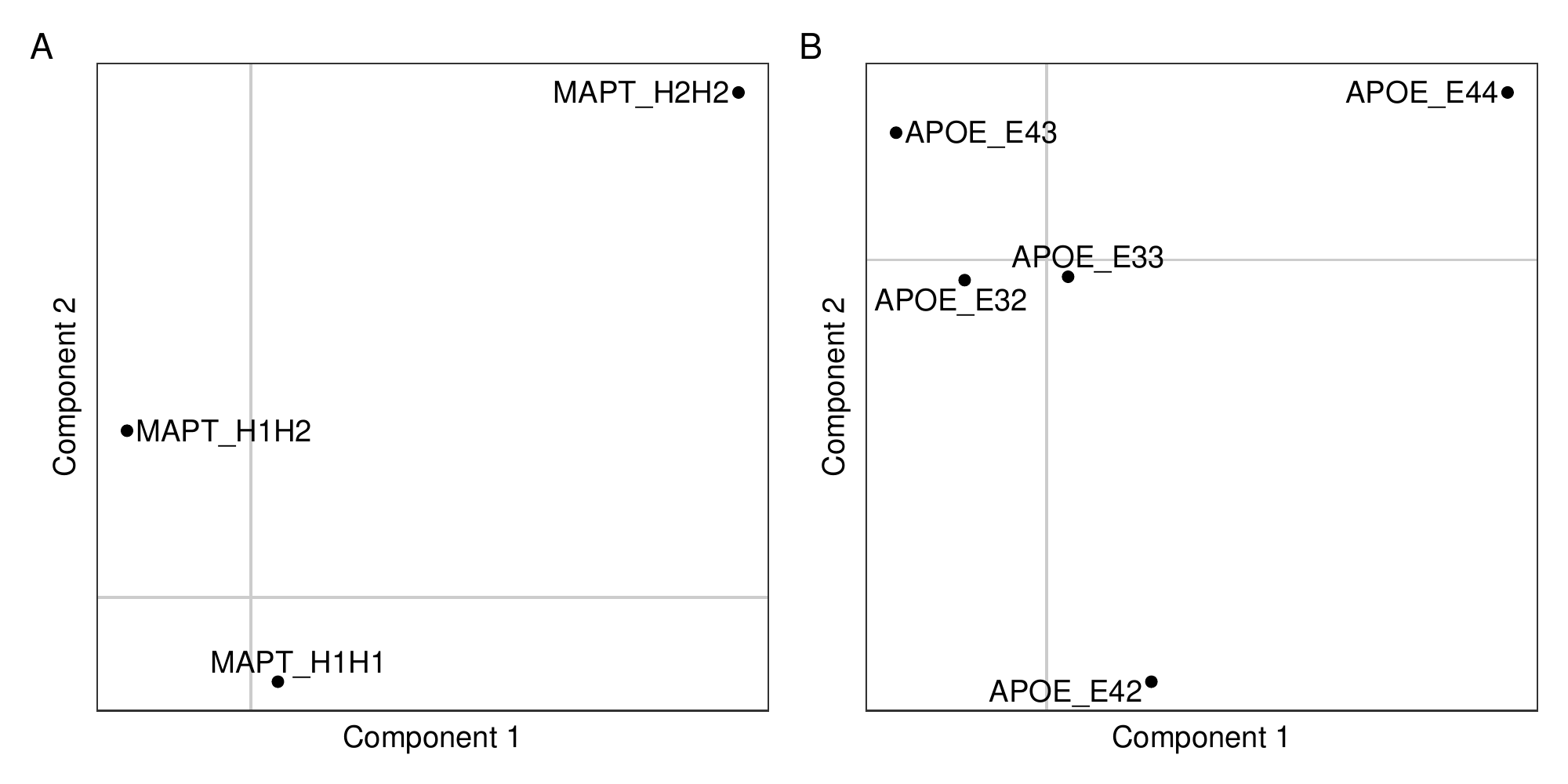} 

}

\caption{\label{fig:ca} Row and column component scores for Correspondence Analysis (CA). Panel A shows the row component scores, where panel B shows the column component scores.}\label{fig:ca_vis}
\end{figure}
\end{CodeChunk}

\begin{CodeChunk}
\begin{table}

\caption{\label{tab:mca_data_prep}\label{tab:disj} Illustration of complete disjunctive coding (a.k.a. dummy coding, one-hot encoding) where each level of a categorical variable is represented. A value of '1' indicates the presence of that level for that row ('0' otherwise).}
\centering
\begin{tabular}[t]{lrrrrr}
\toprule
  & MAPTH1H1 & MAPTH1H2 & MAPTH2H2 & APOEE32 & APOEE33\\
\midrule
OND01\_SYN\_0001 & 0 & 0 & 1 & 0 & 0\\
OND01\_SYN\_0011 & 1 & 0 & 0 & 0 & 1\\
OND01\_SYN\_0021 & 0 & 1 & 0 & 0 & 1\\
OND01\_SYN\_0041 & 1 & 0 & 0 & 0 & 1\\
OND01\_SYN\_0083 & 1 & 0 & 0 & 0 & 1\\
\bottomrule
\end{tabular}
\end{table}

\end{CodeChunk}

Next I show MCA, which generalizes standard CA to multiple categorical
variables. Here we use four variables: MAPT and ApoE like before, and
now also include sex, and a clinical measure with a few ordinal levels
(but here we treat those as levels of a categorical variable). The
computation for MCA is exactly the same as CA, but now the data are
disjunctive (see Table \ref{tab:disj}). However, MCA is more akin to PCA
with observations on the rows and measures on the columns, as opposed to
standard CA which has measures on both the rows and columns. The results
of MCA are shown in Figure \ref{fig:mca}, where Figure \ref{fig:mca}a
shows the component scores for the rows (observations) and Figure
\ref{fig:mca}b shows the columns (measures, some of which are seen in
Table \ref{tab:disj}).

\begin{CodeChunk}
\begin{CodeInput}
R> observed_matrix <- disjunctive_data / sum(disjunctive_data)
R> row_probabilities <- rowSums(observed_matrix)
R> col_probabilities <- colSums(observed_matrix)
R> expected_matrix <- row_probabilities %o% col_probabilities
R> deviations_matrix <- observed_matrix - expected_matrix
R> 
R> mca_gsvd <- gsvd( deviations_matrix, 
+                   LW = 1/row_probabilities,
+                   RW = 1/col_probabilities)
\end{CodeInput}
\end{CodeChunk}

\begin{CodeChunk}
\begin{figure}

{\centering \includegraphics{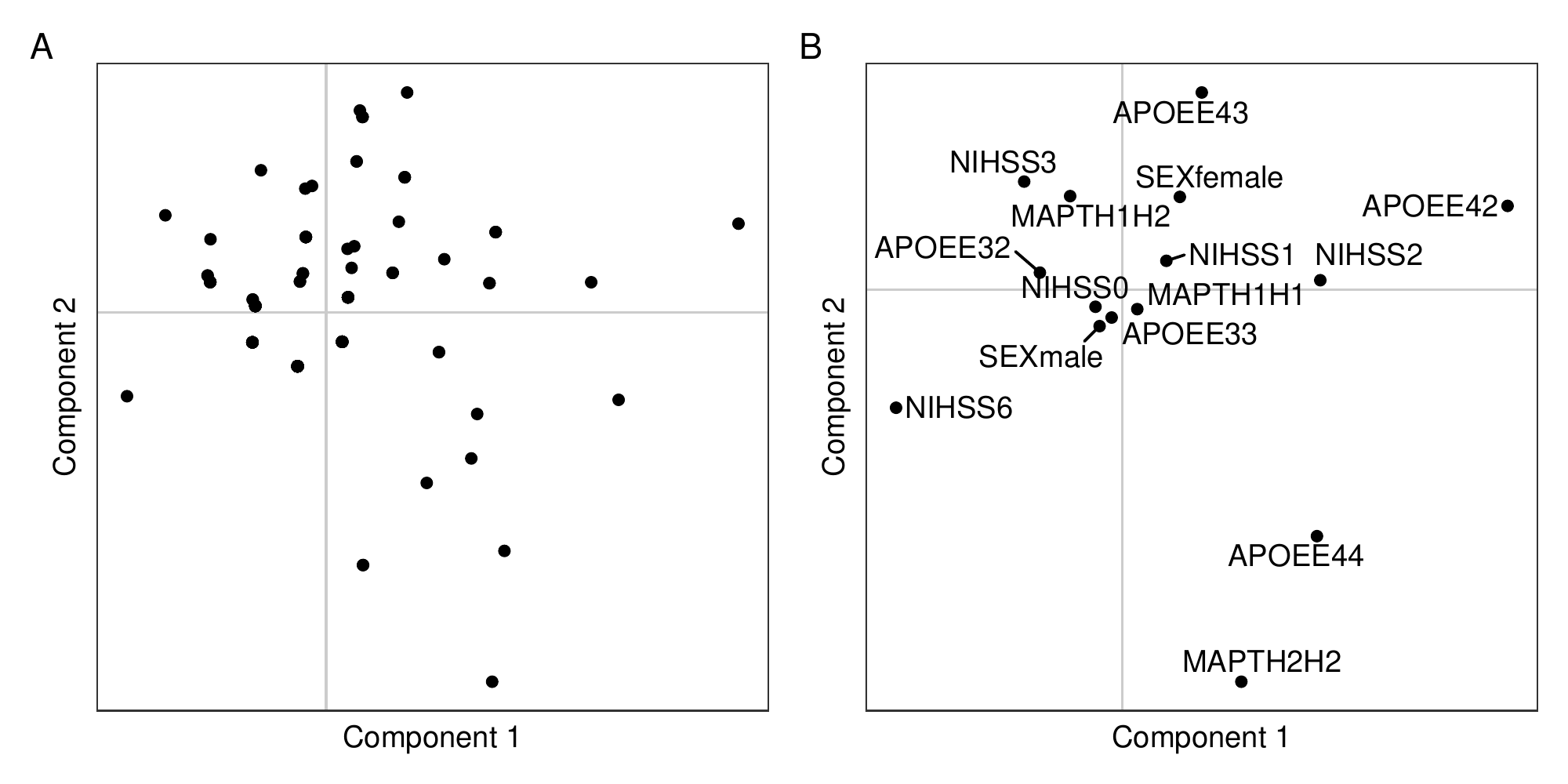} 

}

\caption{\label{fig:mca} Row and column component scores from Multiple Correspondence Analysis (MCA). Here, the row component scores (panel A) are observations, where the column component scores (panel B) are levels of categorical variables.}\label{fig:mca_vis}
\end{figure}
\end{CodeChunk}

Up until this point, all constraints matrices have been diagonal
matrices. The next example serves two purposes: (1) use a
well-established method that has more sophisticated constraints than
simple vectors (or diagonal matrices), and (2) highlight a particular
ridge regularization approach to show how the GSVD easily incorporates
such techniques. Takane \citep{takane_regularized_2006} introduced a
variation of MCA called regularized MCA (RMCA) that uses a ridge-like
approach. The ridge-like approach requires a cross product matrix for
the row (observation) constraints, and a block-diagonal matrix for the
column (measures) constraints. The next block of code shows how the
procedure works through the GSVD. When the regularization
parameter---\(\omega\)---is 0, then this is the same as standard MCA
(within a scaling factor). Also note that for each iteration of RMCA, we
now make use of the \code{k} parameter where \code{gsvd(..., k = 2)}, so
that we only return a subset of possible results. All sets of vectors
and scores are just the first two components, with the exception of
\code{\$d_full} and \code{\$l_full} which return the full set of
singular and eigenvalues, respectively. Note the small changes in the
output object that indicate how many full (possible) components exist,
and also how many were returned (\code{k=2}).

\begin{CodeChunk}
\begin{CodeInput}
R> omegas <- c(0, 1, 2, 3, 4, 5, 10, 25, 50)
R> rmca_results <- vector("list",length(omegas))
R> 
R> centered_data <- scale(disjunctive_data,scale=F)
R> projmat <- t(centered_data) %*% 
+   MASS::ginv( tcrossprod(centered_data) ) %*% 
+   centered_data
R>   
R> for(i in 1:length(omegas)){
+   
+   LW <- diag(nrow(centered_data)) + 
+     (omegas[i] * MASS::ginv(tcrossprod(centered_data)))
+   RW <- diag(colSums(disjunctive_data)) + (omegas[i] * projmat)
+   invRW <- t(MASS::ginv(RW))
+   
+   rownames(LW) <- colnames(LW) <- rownames(centered_data)
+   rownames(invRW) <- rownames(RW)
+   colnames(invRW) <- colnames(RW)
+   
+   rmca_results[[i]] <- gsvd(centered_data, LW = LW, RW = invRW, k = 2)
+   
+ }
\end{CodeInput}
\end{CodeChunk}

\begin{CodeChunk}
\begin{CodeInput}
R> rmca_results[[1]]
\end{CodeInput}
\begin{CodeOutput}
**GSVD package object of class type 'gsvd'.**

gsvd() was performed on a matrix with 138 rows and 15 columns
Number of components = 11.
Number of retained components = 2.

The 'gsvd' object contains:                                                                
 $d_full Full set of singular values                            
 $l_full Full set of eigen values                               
 $d      Retained set of singular values (k)                    
 $l      Retained set of eigen values (k)                       
 $u      Left singular vectors (for rows of DAT)                
 $v      Right singular vectors (for columns of DAT)            
 $p      Left generalized singular vectors (for rows of DAT)    
 $q      Right generalized singular vectors (for columns of DAT)
 $fi     Left component scores (for rows of DAT)                
 $fj     Right component scores (for columns of DAT)            
\end{CodeOutput}
\end{CodeChunk}

So what does this regularization do? In Figure \ref{fig:rmca_v_mca} we
show a partial view of how this regularization works. In Figures
\ref{fig:rmca_v_mca}a and b we see the scree plots, which is the
explained variance (eigenvalues) per component. In Figure
\ref{fig:rmca_v_mca}c and d we see the component scores for the columns.
Figure \ref{fig:rmca_v_mca}a and c show the standard MCA where Figure
\ref{fig:rmca_v_mca}b and d show each iteration of the RMCA, where the
\(\omega\) parameter is used to color the eigenvalues and component
scores on a gradient scale (via
\code{ggplot::scale_color_gradient(...,trans="log1p")}). Figure
\ref{fig:rmca_v_mca} shows that as the regularization parameter
increases, we see that explained variance for the higher dimensions
\emph{decreases}, and that in general, the component scores approach
zero. Figure \ref{fig:rmca_v_mca}d in particular also shows us that
stable measures do not change much under regularization (e.g.,
\texttt{SEXfemale}) where as rare and relatively unstable measures
approach zero and see substantial changes in their component scores
(e.g., \texttt{MAPTH2H2} or \texttt{APOEE42}).

The CA-based approaches make use of each element in the GSVD. The
implementations here show that standard CA and MCA use
vectors---algebraically these are diagonal matrices---where RMCA
requires more sophisticated matrices than simple diagonal matrices. RMCA
illustrates two key concepts: (1) with appropriate sets of constraints,
the GSVD---and all of the generalized methods presented here---provides
an easy pathway to ridge regularization, and (2) constraints matrices
are not always simple diagonal matrices; an important point expanded
upon in the next section.

\begin{CodeChunk}
\begin{figure}

{\centering \includegraphics{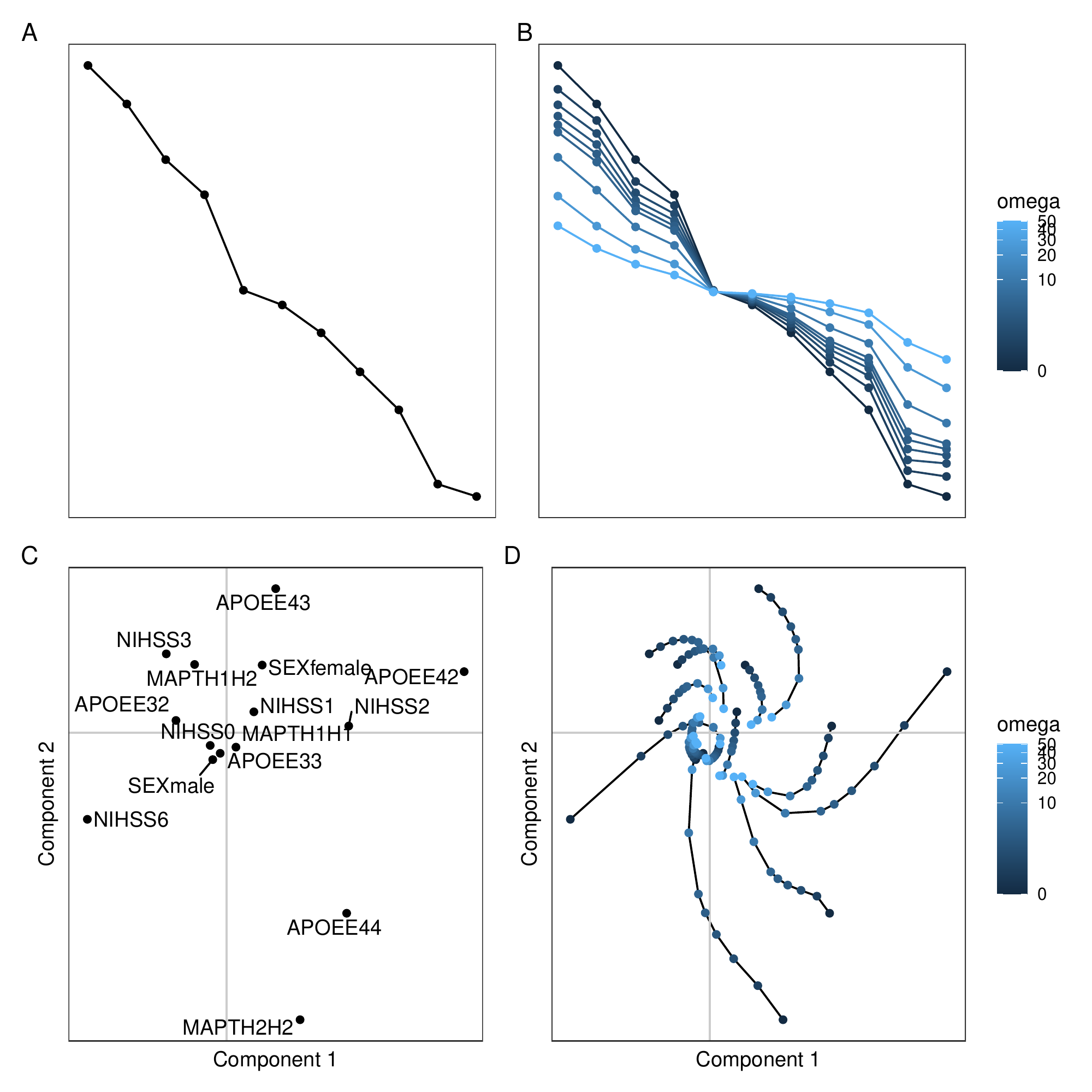} 

}

\caption{\label{fig:rmca_v_mca} MCA (panels A and C) vs. Regularized MCA (panels B and D). Panels A and B show scree plots, which reflect the amount of explained variance per component (and thus decreases as components increase). RMCA shows multiple overlaid screes because there exists one version for each run with $\omega$. Panels C and D show the column component scores (levels of categorical variables) in the plain MCA and as the scores change through the iterations of RMCA. Only the first two components are shown, with the first component on the horizontal axis, and the second component on the vertical axis. Note that the initial RMCA ($\omega = 0$) is equivalent to MCA.}\label{fig:unnamed-chunk-1}
\end{figure}
\end{CodeChunk}

\hypertarget{partial-least-squares-reduced-rank-regression-and-canonical-correlation}{%
\subsection{Partial least squares, reduced rank regression, and
canonical
correlation}\label{partial-least-squares-reduced-rank-regression-and-canonical-correlation}}

So far, we have seen various examples of the GEVD and GSVD. The GEVD and
GSVD are well-established concepts. Recently, I introduced an extension
of the GEVD and GSVD concept to ``two (data) table techniques'', and
called it this new extension the ``GPLSSVD''
\citep{beaton_generalization_2019}, which is short for the ``generalized
partial least squares-singular value decomposition''. The GPLSSVD
provides the same base concepts as the GEVD or GSVD, but specifically
for scenarios wherein there are two data matrices. The most common
two-table techniques are canonical correlation analysis (CCA), reduced
rank regression (RRR; sometimes called redundancy analysis, RDA), and a
specific form of partial least squares (PLS) called the PLSSVD,
``partial least squares correlation'' (PLSC), and a variety of other
names. This particular form of PLS traces its history to Tucker's
interbattery factor analysis.

The GPLSSVD requires at minimum two data matrices, and allows for the
inclusion of positive semi-definite constraints matrices for the rows
and columns of both data matrices. Here, I illustrate four methods that
make use of the GPLSSVD. All of these examples emphasize the expression
of latent variables, though do recall that at their core, these methods
make use of the SVD. Each method makes use of different sets of
constraints: (1) PLS-SVD/PLSC which uses no constraints, (2) RRR/RDA
which uses one set of constraints, (3) CCA which uses two sets of
constraints, and finally (4) partial least squares-correspondence
analysis (PLS-CA; \citet{beaton_generalization_2019};
\citet{beaton_partial_2016}) which makes use of all four sets of
constraints. The first three methods---PLS, RRR, and CCA---are methods
used to analyze numeric, or data that are generally assumed to be
continuous values. The final method---PLS-CA---was initially designed
for analysis of two data sets that are categorical, but easily extends
to a variety of data types \citep{beaton_generalization_2019}. With that
in mind, the first three examples will all make use of exactly the same
data sets. This helps highlight the similarities and differences between
PLS, RRR, and CCA, and shows how the same data can be viewed in multiple
ways by imposing various constraints.

In the PLS/RRR/CCA examples, we have two data matrices of generally
continuous data. One matrix contains age, total gray matter volume (in
percentage of total intracranial volume), and total white matter volume
(in percentage of total intracranial volume). The other matrix contains
the cognitive tasks seen in the PCA and MDS examples. Here, the
respective GPLSSVD models for each method are:

\begin{itemize}
\item
  PLS:
  \(\mathrm{GPLSSVD(}{\bf I}_{N}, {\bf X},{\bf I}_{I},{\bf I}_{N},{\bf Y},{\bf I}_{J}\mathrm{)}\)
\item
  RRR:
  \(\mathrm{GPLSSVD(}{\bf I}_{N}, {\bf X}, ({\bf X}^{T}{\bf X})^{-1},{\bf I}_{N},{\bf Y},{\bf I}_{J}\mathrm{)}\)
\item
  CCA:
  \(\mathrm{GPLSSVD(}{\bf I}_{N}, {\bf X}, ({\bf X}^{T}{\bf X})^{-1},{\bf I}_{N},{\bf Y},({\bf Y}^{T}{\bf Y})^{-1}\mathrm{)}\)
\end{itemize}

Before an illustration of these techniques, it is worth noting that PLS,
RRR, and CCA are multivariate extensions of univariate concepts. PLS
emphasizes the covariance between \({\bf X}\) and \({\bf Y}\), RRR
emphasizes the least squares fit (i.e., regression) of \({\bf Y}\) onto
space defined by \({\bf X}\), and CCA emphasizes the correlation between
\({\bf X}\) and \({\bf Y}\). For the first illustrations of PLS, RRR,
and CCA, I use column-wise centered and scaled \({\bf X}\) and
\({\bf Y}\) matrices.

\begin{CodeChunk}
\begin{CodeInput}
R> X <- synthetic_ONDRI[,
+                      c("TMT_A_sec", "TMT_B_sec",
+                        "Stroop_color_sec","Stroop_word_sec",
+                        "Stroop_inhibit_sec","Stroop_switch_sec")]
R> Y <- synthetic_ONDRI[,c("AGE","NAGM_PERCENT","NAWM_PERCENT")]
R> 
R> scaled_X <- scale(X, center = T, scale = T)
R> scaled_Y <- scale(Y, center = T, scale = T)
R> 
R> 
R> pls_gplssvd <- gplssvd(scaled_X, scaled_Y)
R> 
R> rrr_gplssvd <- gplssvd(scaled_X, scaled_Y, 
+                        XRW = MASS::ginv(crossprod(scaled_X)))
R> 
R> cca_gplssvd <- gplssvd(scaled_X, scaled_Y, 
+                        XRW = MASS::ginv(crossprod(scaled_X)), 
+                        YRW = MASS::ginv(crossprod(scaled_Y)))
\end{CodeInput}
\end{CodeChunk}

All three approaches provide the same types of outputs: singular and
eigenvalues, latent variable scores (for the participants), standard and
generalized singular vectors, and component scores. The output object is
identical for all three approaches because they all make use of
\code{gplssvd()}, so let's only look at one of the objects,
\texttt{cca\_gplssvd}. Like both the \code{geigen()} and \code{gsvd()}
outputs, we have many common objects (e.g., vectors, scores,
eigenvalues). However \code{gplssvd()} also provides the latent variable
scores---\code{\$lx} and \code{\$ly}---which are row scores---for
\({\bf X}\) and \({\bf Y}\), respectively---with respect to the singular
vectors. For reference, see also Tables \ref{tab:arguments} and
\ref{tab:values}.

\begin{CodeChunk}
\begin{CodeInput}
R> cca_gplssvd
\end{CodeInput}
\begin{CodeOutput}
**GSVD package object of class type 'gplssvd'.**

gplssvd() was performed on an X matrix with 138 rows and 6 columns and a Y matrix with 138 rows and 3 columns
Number of total components = 3.
Number of retained components = 3.

The 'gplssvd' object contains:                                                              
 $d_full Full set of singular values                          
 $l_full Full set of eigen values                             
 $d      Retained set of singular values (k)                  
 $l      Retained set of eigen values (k)                     
 $u      Left singular vectors (for columns of X)             
 $v      Right singular vectors (for columns of Y)            
 $p      Left generalized singular vectors (for columns of X) 
 $q      Right generalized singular vectors (for columns of Y)
 $fi     Left component scores (for columns of X)             
 $fj     Right component scores (for columns of Y)            
 $lx     Left (X) latent variable scores (for rows of X)      
 $ly     Right (Y) latent variable scores (for rows of Y)     
\end{CodeOutput}
\end{CodeChunk}

Before visualizing the results of these three approaches, let's first
establish the connections between framing the methods via the GPLSSVD
and the more typical ways. Let's do so by pointing out the relationship
between \code{base::cancor} and the results in \code{cca_gplssvd}. From
\code{cca_gplssvd}, the canonical correlations are the singular values.
Recall also that the optimization when framed through the GPLSSVD is
that the singular values are also
\(\mathrm{diag\{}{\bf L}_{\bf X}^{T}{\bf L}_{\bf Y}\mathrm{\}}\), so
\code{diag( t(cca_gplssvd\$lx) \%*\% cca_gplssvd\$ly )} are also the
canonical correlations. So from the perspective of GPLSSVD, the
relationship between the latent variables also maximize for the
canonical correlations. Furthermore, the coefficients (per set of
variables) from \code{base::cancor(scaled_X, scaled_Y)} are equal to
\({\bf W}_{\bf X}{\bf P}\) and \({\bf W}_{\bf Y}{\bf Q}\) in the GPLSSVD
model or alternatively also \({\bf F}_{I}{\bf \Delta}^{-1}\) or
\({\bf F}_{J}{\bf \Delta}^{-1}\). The primary difference between the
GPLSSVD approach and various implementations of canonical correlation
(including \code{base::cancor()}) is that all sets of scores from the
GPLSSVD are limited to the rank of the decomposed matrix. That is, if
\({\bf X}\) has fewer variables than \({\bf Y}\), then all sets of
scores associated with \({\bf Y}\) will be only span the vectors
determined by the rank of the relationship between those matrices.
Likewise, in reduced rank regression (a.k.a. redundancy analysis) there
exist \(\beta\) coefficients which are expressed through the GPLSSVD
model above as \({\bf P}{\bf \Delta}\) or
\code{rrr_gplssvd\$p \%*\% diag(rrr_gplssvd\$d)}. Finally, it is worth
noting that the above approaches to RRR and CCA are not the only ways to
frame these techniques through the GPLSSVD. There are multiple and, in
some cases, computationally simpler alternatives; however those
approaches are conceptually more abstract than how I framed them here.
Plus, the \pkg{GSVD} framework highlights how these techniques stem from
the same core methods.

Let's now visually compare the results from the GPLSSVD for PLS, RRR,
and CCA. First let's look at the latent variable scores, which expresses
the maximization in GPLSSVD as scores for the rows (observations) of
each data matrix.

\begin{CodeChunk}
\begin{figure}

{\centering \includegraphics{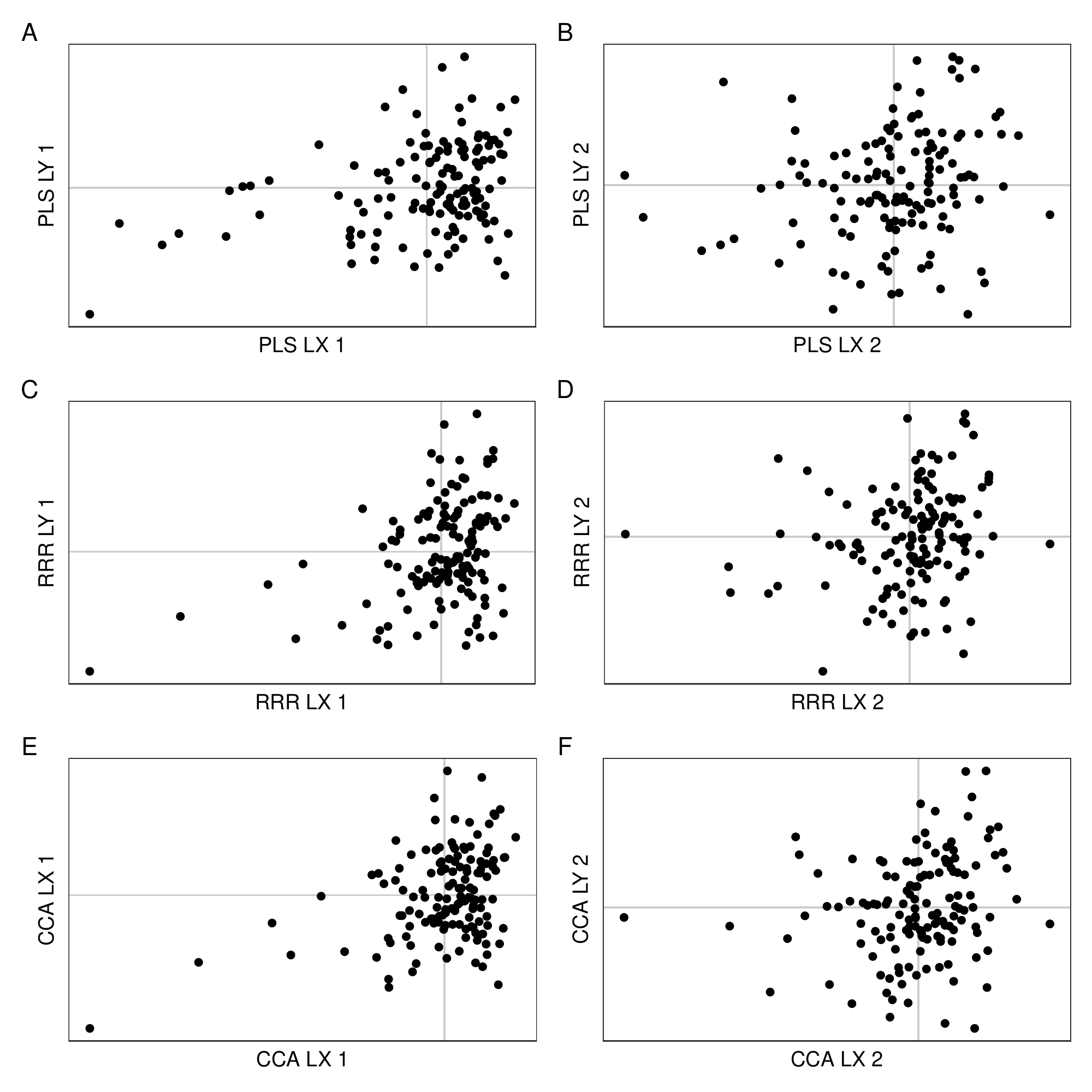} 

}

\caption{\label{fig:tt_ls}Latent variable scores for PLS, RRR, and CCA where with data matrices are column-wise centered and scaled. The latent variable scores for the rows of each data matrix, with the row items of $\bf X$ on the horizontal axes and with the row items of $\bf Y$ on the vertical axes. Panels A, C, and E show the first latent variable (X vs. Y) and panels B, D, and E show the second latent variable (X vs. Y). Panels A and B show PLS, panels C and D show RRR, and panels E and F show CCA.}\label{fig:two_tables_vis_ls}
\end{figure}
\end{CodeChunk}

Figure \ref{fig:tt_ls} shows the latent variable scores for all three of
the analyses. Latent variable scores are the expression of the rows of
each data matrix wth respect to the components (a.k.a. latent
variables). Each data matrix---i.e., \({\bf X}\) and \({\bf Y}\)---has a
set of latent variable scores per component. Typically in PLS, the
latent variables for each matrix are plotted against one another, that
is, the first column of \({\bf L}_{\bf X}\) vs.~the first column of
\({\bf L}_{\bf Y}\). Plotting the latent variables reflects what the
GPLSSVD maximizes (i.e., \({\bf L}_{\bf X}^{T}{\bf L}_{\bf Y}\)). So we
show these latent variables scores for all of the GPLSSVD approaches
(PLS, RRR, and CCA).

Figure \ref{fig:tt_ls} shows that the results are highly similar for the
latent variable scores across approaches (a point I revisit soon). That
is, these techniques show an approximately equivalent perspective of how
the rows (observations) express themselves with respect to the latent
variables. What about the variables of each data set? Figure
\ref{fig:tt_fs} shows the component scores from these same analyses.
Figure \ref{fig:tt_fs} shows some clear differences in how variables are
related across the techniques---in particular Figure \ref{fig:tt_fs}a
from PLS. The similarities and differences in Figure \ref{fig:tt_fs}
highlight that while these techniques are very related, they do not
necessarily express the relationships between \({\bf X}\) and
\({\bf Y}\) in the same way.

\begin{CodeChunk}
\begin{figure}

{\centering \includegraphics{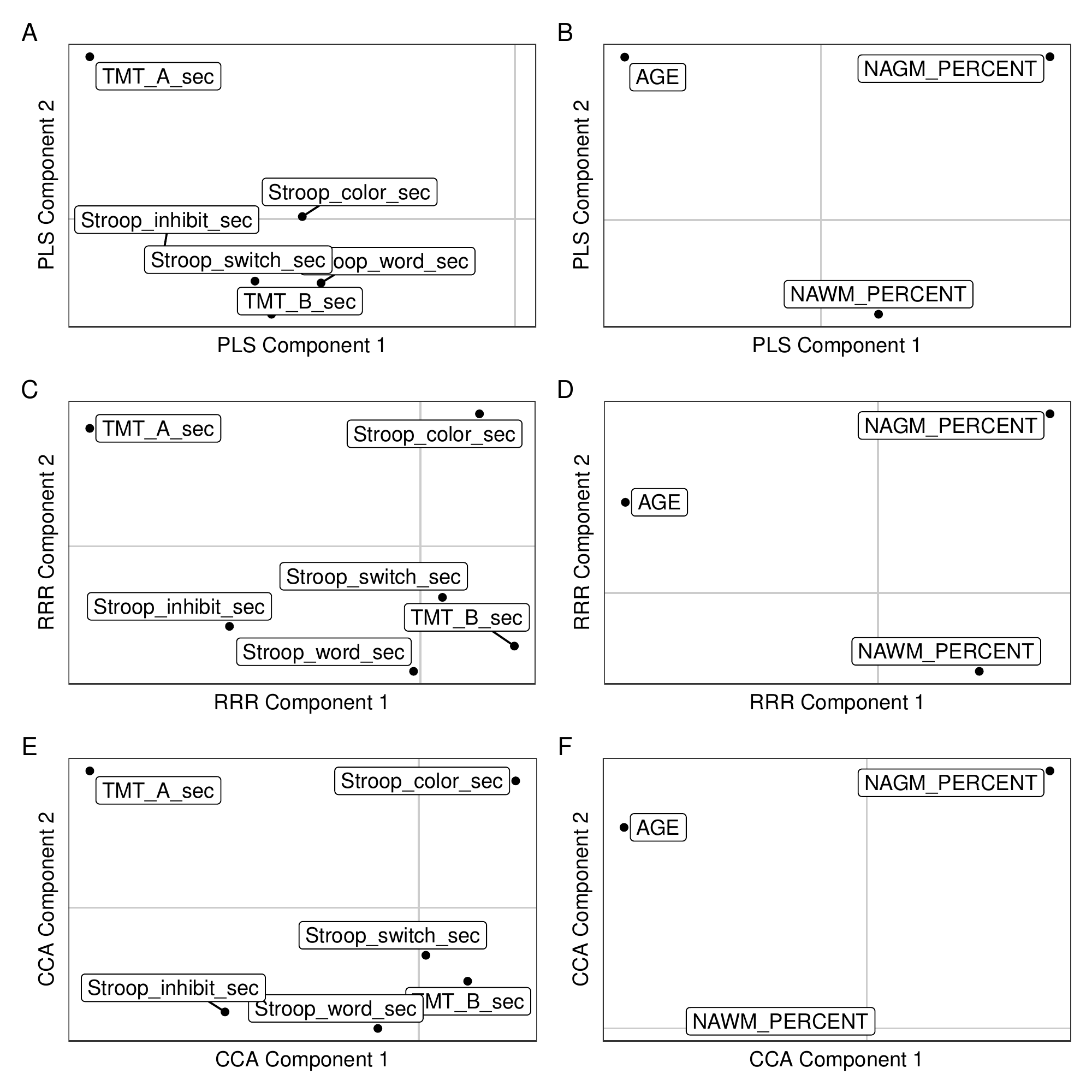} 

}

\caption{\label{fig:tt_fs}Component scores for PLS, RRR, and CCA where with data matrices are column-wise centered and scaled. Panels A, C, and E show the variables from the $\bf X$ matrix, where panels B, D, and E show the variables from the $\bf Y$ matrix. The horizontal axes are the first component, where the vertical axes are the second component. Panels A and B show PLS, panels C and D show RRR, and panels E and F show CCA. }\label{fig:two_tables_vis_fs}
\end{figure}
\end{CodeChunk}

Recall that for these analyses, GPLSSVD optimizes for the maximum common
information between the two data sets---with respect to their
constraints---where
\({\bf L}_{\bf X}^{T}{\bf L}_{\bf Y} = {\bf \Delta} = {\bf P}^{T}{\bf W}_{\bf X}[({\bf M}_{\bf X}^{\frac{1}{2}}{\bf X})^{T}({\bf M}_{\bf Y}^{\frac{1}{2}}{\bf Y})]{\bf W}_{\bf Y}{\bf Q}\).
Now let's look at these same approaches but where the data matrices are
column-wise centered but not scaled. Figures \ref{fig:tt2ls} and
\ref{fig:tt2fs} show the latent variable scores and the component scores
where the data are only column-wise centered. Now the analyses show much
more apparent differences. The differences are because the data
transformations in conjunction with the constraints, each play an
important role in the results, and thus critical roles for the
interpretation of those results.

\begin{CodeChunk}
\begin{figure}

{\centering \includegraphics{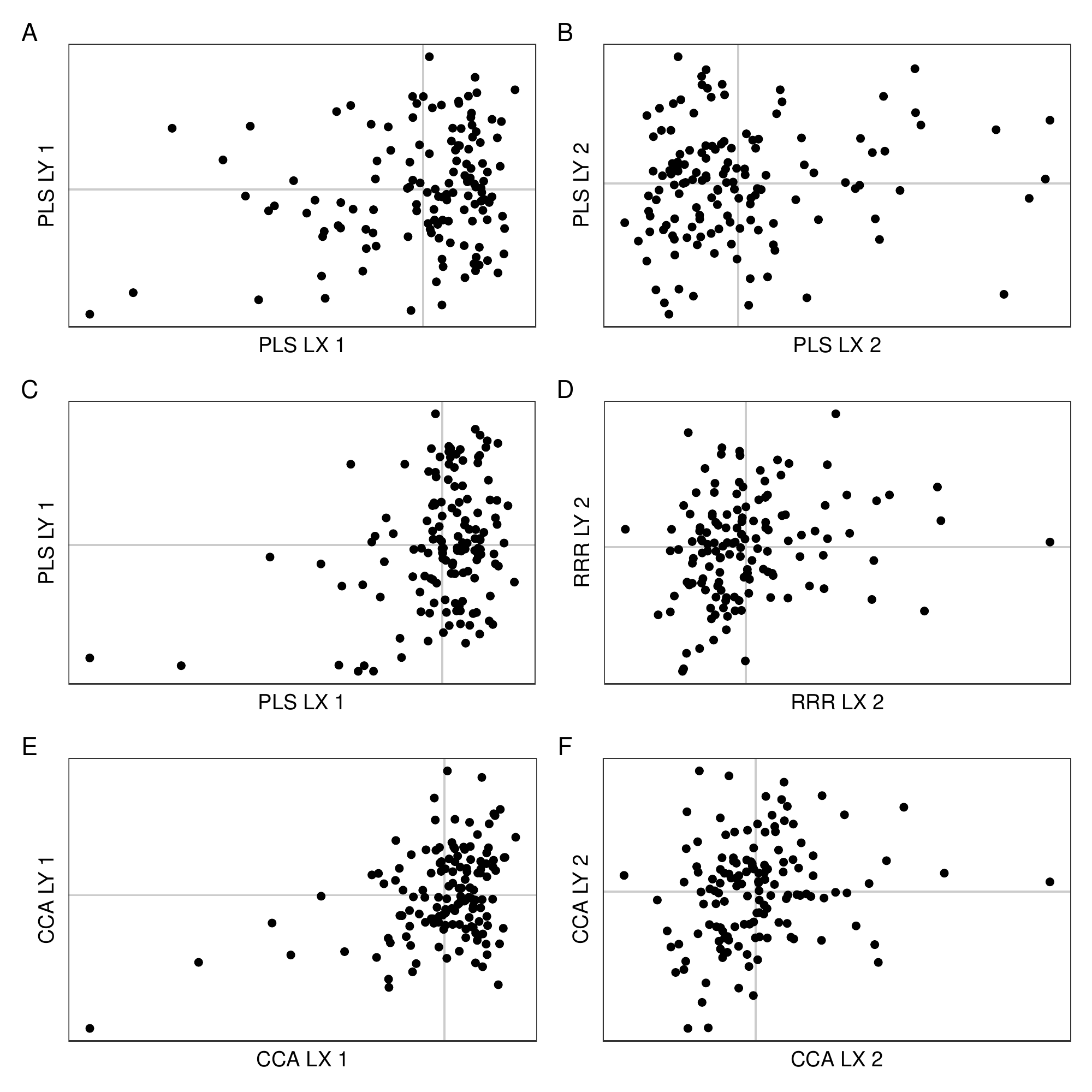} 

}

\caption{\label{fig:tt2ls}Latent variable scores for PLS, RRR, and CCA, but now with data only column-wise centered. The latent variable scores for the rows of each data matrix, with the row items of $\bf X$ on the horizontal axes and with the row items of $\bf Y$ on the vertical axes. Panels A, C, and E show the first latent variable (X vs. Y) and panels B, D, and E show the second latent variable (X vs. Y). Panels A and B show PLS, panels C and D show RRR, and panels E and F show CCA.}\label{fig:two_table2_vis_ls}
\end{figure}
\end{CodeChunk}

\begin{CodeChunk}
\begin{figure}

{\centering \includegraphics{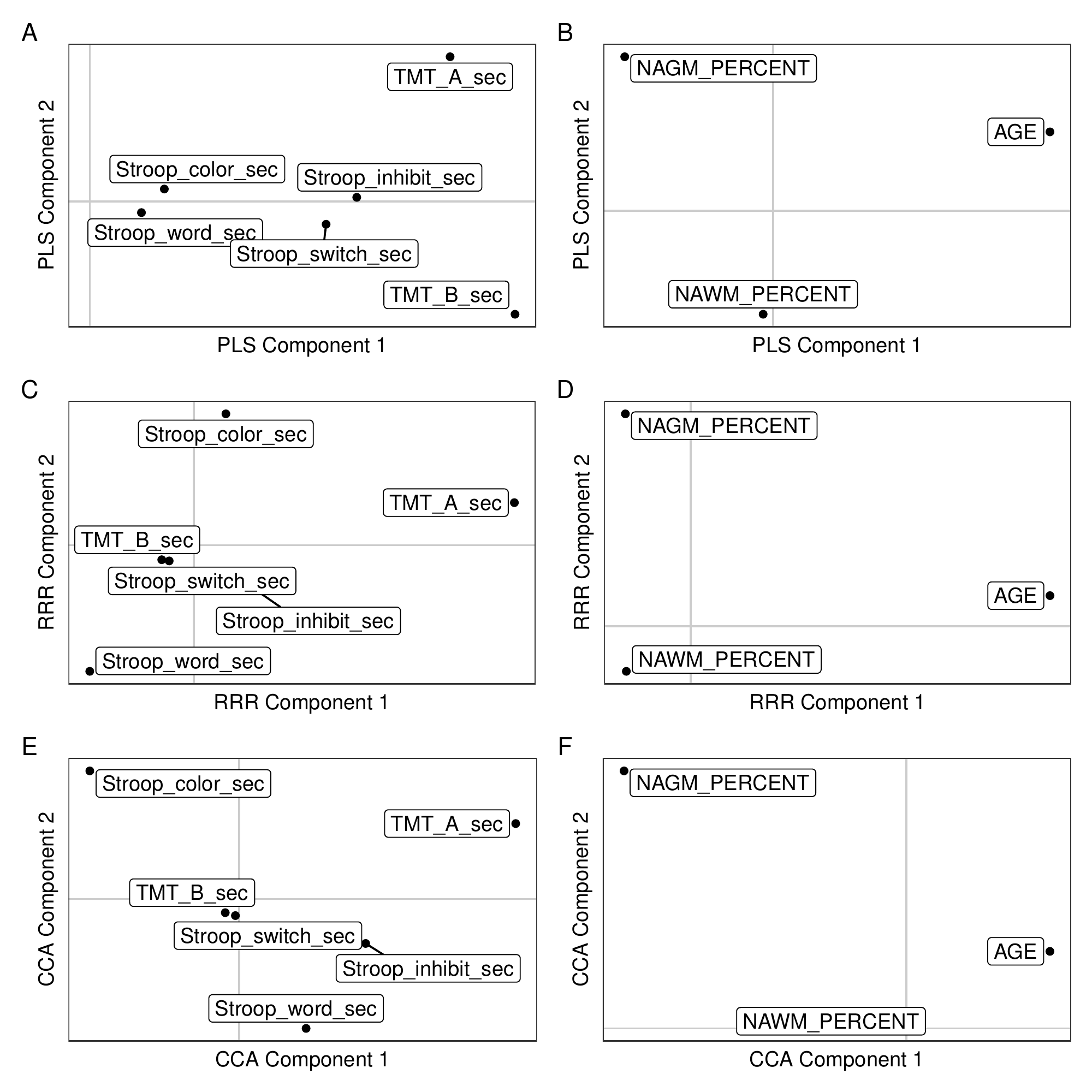} 

}

\caption{\label{fig:tt2fs}Component scores for PLS, RRR, and CCA, but now with the data only column-wise centered. Panels A, C, and E show the variables from the $\bf X$ matrix, where panels B, D, and E show the variables from the $\bf Y$ matrix. The horizontal axes are the first component, where the vertical axes are the second compnent. Panels A and B show PLS, panels C and D show RRR, and panels E and F show CCA. }\label{fig:two_table2_vis_fs}
\end{figure}
\end{CodeChunk}

Finally, let's look at a GPLSSVD approach that makes use of all data and
constraint matrices: PLS-CA. PLS-CA was initially designed as a PLS
approach for categorical data \citep{beaton_partial_2016} but can
accommodate virtually any data type \citep{beaton_generalization_2019}.
Here, let's just focus on the problem of two categorical tables. The
data for PLS-CA are processed in the same way they are for MCA, except
now there are two tables. For the PLS-CA example, we have genetics in
one table and age plus a clinical measure in the other table. All of
these measures are categorical or ordinal. In this example and for
simplicity, the ordinal data are treated as distinct levels instead of
ordered levels.

\begin{CodeChunk}
\begin{CodeInput}
R> plsca_gplssvd <- gplssvd( 
+                   X = deviations_matrix_X,
+                   Y = deviations_matrix_Y,
+                   XLW = 1/row_probabilities_X,
+                   YLW = 1/row_probabilities_Y,
+                   XRW = 1/col_probabilities_X,
+                   YRW = 1/col_probabilities_Y
+                   )
\end{CodeInput}
\end{CodeChunk}

Like the other GPLSSVD techniques, PLS-CA also produces latent variable
scores. But we forgo visualizing those in favor of highlighting the
singular vectors and scores. That's because PLS-CA is one particular
method where we can much more clearly see the differences between the
singular vectors, the generalized singular vectors, and the component
scores. Figure \ref{fig:plsca} shows all sets of these scores. Figures
\ref{fig:plsca}a, c, and e show the scores associated with the columns
of one data matrix, where as Figure \ref{fig:plsca}b, d, and f show the
scores for the other data matrix. Figures \ref{fig:plsca}a and b show
the singular vectors, Figures \ref{fig:plsca}c and d show the
\emph{generalized} singular vectors, and Figures \ref{fig:plsca}e and f
show the component scores. Figures \ref{fig:plsca} shows component 1 on
the horizontal axis with component 2 on the vertical axis for all
subfigures.

\begin{CodeChunk}
\begin{figure}

{\centering \includegraphics{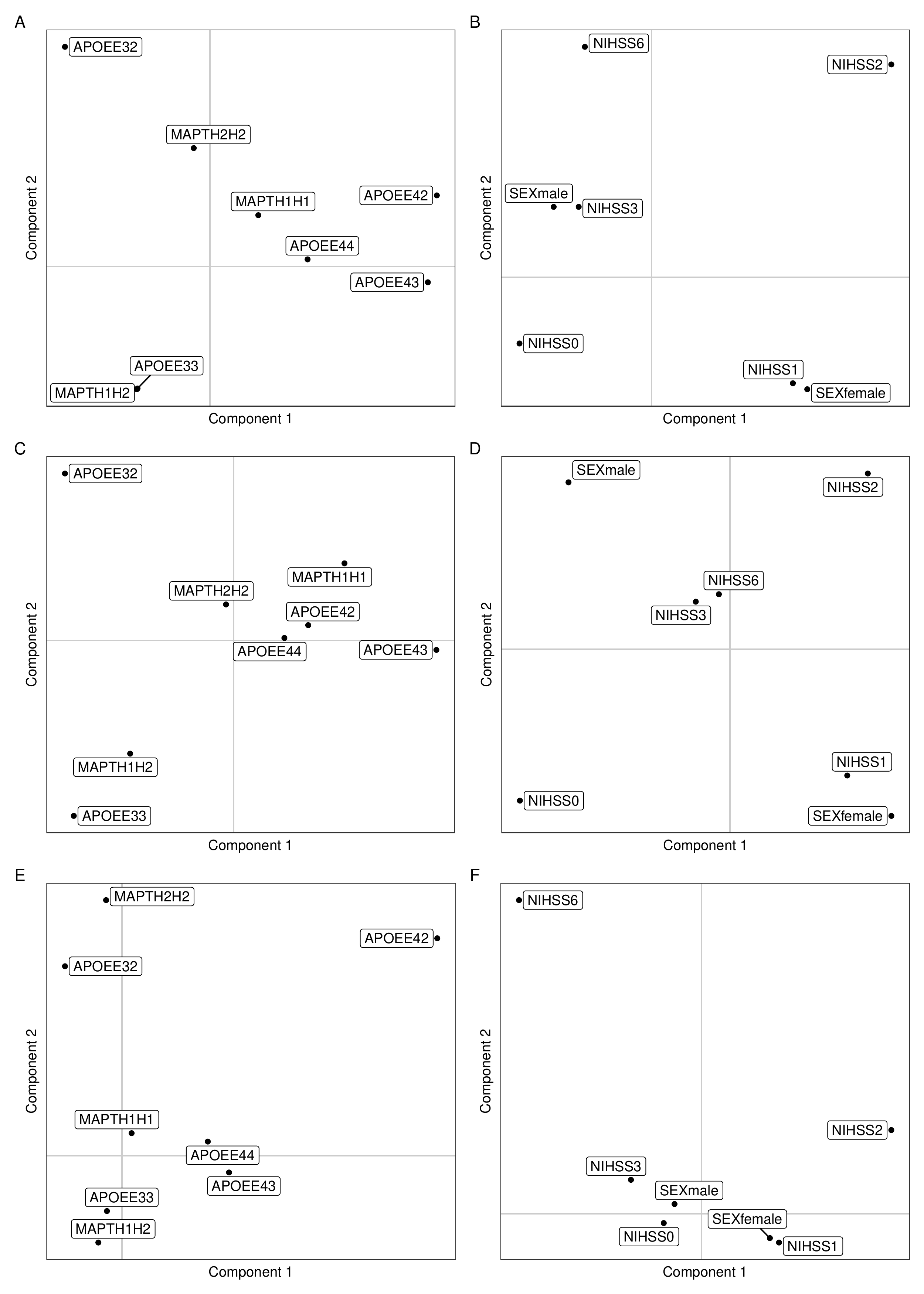} 

}

\caption{\label{fig:plsca}Singular vectors, generalized singular vectors, and component scores from partial least squares-correspondence analysis (PLS-CA). The horizontal axes are the first component, and the vertical axes are the second component. Panels A and B show the singular vectors, panels C and D show the generalized singular vectors, and Panels E and D show the component scores. Panels A, C, and E show the columns from the $\bf X$ matrix, where panels B, D, and F show the columns from the $\bf Y$ matrix. Each visual helps reflect the types and amount of information we can obtain from GPLSSVD methods that make use of various sets of weights.}\label{fig:plsca_vis}
\end{figure}
\end{CodeChunk}

\hypertarget{final-notes}{%
\subsection{Final notes}\label{final-notes}}

There are several influential factors that drive the core principles
that motivated the design and implementation of the \pkg{GSVD} package.
These influences span the ``French way'' of data analyses
\citep{holmes_discussion_2017, holmes_multivariate_2008}, the
relationships between various decompositions
\citep{takane_relationships_2003}. Furthermore, many of the most common
multivariate techniques perform these non-arbitrary transformations via
(positive semi-definitive) metrics used as constraints (see Takane GEVD
\& GSVD). The metrics used for constraints include, for examples,
\(\chi^2\) distances for Correspondence Analysis, Mahalanobis distances
for CCA and related methods. At its core, the \pkg{GSVD} package
provides the three primary bases---\code{geigen()}, \code{gsvd()}, and
\code{gplssvd()}---of the most common multivariate techniques and
numerous variants of those techniques. The package provides a radically
simplified way to approach these methods: users generally need only to
know what their constraints and (transformed) data should be, instead
about how to perform the necessary steps (e.g., matrix multiplication,
square roots of matrices).

The \code{eigen()} and \code{svd()} functions---and many of their
analogs and extensions---provide a very simplified set of outputs.
However, with respect to core analysis and interpretation principles,
the \pkg{GSVD} package provides comprehensive output common to the
numerous eigen-based multivariate techniques. The component scores are
examples of important and comprehensive outputs from the \pkg{GSVD}
package. The component scores provide a view of the results with respect
to (1) the constraints, and (2) the length of the components (singular
values). The component scores---and the fact that these are weighted and
``stretched'' versions of the generalized vectors---means that
interpretation and visualization requires caution and care.
\citet{nguyen_ten_2019} provide a comprehensive view of decompositions,
but most importantly make suggestions on how to more appropriately show
and interpret visuals. The visuals presented here did not stretch the
axis according to the proportions of the singular or eigenvalues. That
stretching is important for interpretation, but here many of the visuals
are either simple illustrations of output or for simple comparisons.

Finally, it is worth noting that many techniques can be performed with
any of these decompositions. For examples, PLS, RRR, and CCA could could
be performed with the GSVD (or even with two GEVDs one for each set of
variables). Likewise, one-table techniques (e.g., CA, or PCA) could be
performed as the GSVD or GEVD (as presented earlier here with PCA).
However, the direct interfaces of these methods provide a simpler
conceptual framework for the techniques than some of these alternative
approaches (e.g., CCA as two separate GEVDs).

\hypertarget{discussion}{%
\section{Discussion}\label{discussion}}

The \pkg{GSVD} package provides a core and minimal set of functions
designed around a common framework and nomenclature. More importantly,
\pkg{GSVD} also conceptually unifies a large array of multivariate
techniques. Above, I show only a handful of standard and very common
multivariate approaches, and how they conceptually fit into generalized
decomposition approaches as well as how to perform them with the
\pkg{GSVD} package. The intent and design of \pkg{GSVD} was to provide a
unified core for those analysis approaches. Though users could directly
perform analyses with the \pkg{GSVD} package, there is likely more
benefit to developers and analysts to build analysis methods and
packages on top of \pkg{GSVD}. The primary motivations to build upon
\pkg{GSVD} (i.e., use \pkg{GSVD} as a dependency) are because (1)
\pkg{GSVD} only provides a minimal set of classes and objects for S3
methods, and more importantly (2) numerous multivariate techniques
benefit from additional information; for examples, PCA and CA require
preprocessing steps that are necessary for external or out of sample
data, and thus require such external information (e.g., column centers,
or row weights).

Though only a relatively small set of multivariate techniques were shown
here, the \pkg{GSVD} package was designed for virtually any variation of
some of the methods presented here. Such methods include discriminant
techniques that generally fall under specific instances of PLS, RRR, or
CCA. There are also numerous variations of CA that make use of
different---and sometimes asymmetric---sets of weights
\citep{gimaret1998non}, use power transformations or various centers
\citep{greenacre2009power}, alternate metrics such as the Hellinger
distance \citep{rao1995review}, and numerous other alternatives
\citep[see, e.g.,][]{beh2012genealogy} Even broader families of
techniques that decompose multiple tables also fit within the framework
established here. These multi-table techniques are often ways of
structuring and transforming sets of tables, and then concatenating them
in a way to produce a larger data matrix (with rows and columns). These
multi-table techniques also routinely make use of constraints or weights
imposed on the columns and rows. Some of those methods include, for
examples, constrained PCA \citep{takane2013constrained} and multiple
factor analysis
\citep{abdi_multiple_2013, becue-bertaut_multiple_2008, escofier_multiple_1994}

Finally, the use of returned rank is highly beneficial to statistical
learning and machine learning strategies such as sparsification
techniques such as the penalized matrix decomposition
\citep{witten_penalized_2009} or the constrained SVD
\citep{guillemot_constrained_2019}. Lower rank solutions is even more
common for iterative algorithms like partial least squares regression
(PLSR). The most common implementations of PLSR use a rank 1 SVD
solution, followed by deflation of both the \(\bf X\) and \(\bf Y\)
matrices. To see such PLSR implementations that depend on \pkg{GSVD},
please see another package I have developed that focuses on
two-table/``cross-decomposition'' methods
(\citet{beaton_generalization_2019} see also
\url{https://github.com/derekbeaton/gpls}).

In its current form, the \pkg{GSVD} package is lightweight, has zero
dependencies, and is written in base R. The general goals of the
\pkg{GSVD} package is both to remain lightweight---with minimal yet core
functionality from a user perspective---and as a ``workhorse'' package.
The \pkg{GSVD} package should be the core dependency in an analysis
package, as opposed to being used as the analysis package. Given that, I
plan to provide a substantial overhaul of the \pkg{ExPosition}
\citep{beaton_exposition_2014} family of packages in the (near) future,
and \pkg{GSVD} is the first step towards that overhaul. Though the
\pkg{GSVD} package is intentionally minimal with no dependencies, there
are possible benefits to some dependencies in the future. For example,
there are numerous packages and tools that could make the \pkg{GSVD}
package more memory and computationally efficient. So the future of
\pkg{GSVD} could make use the \pkg{Matrix} \citep{bates_matrix_2019} for
storage (memory) and other tools for faster decompositions themselves.
Specifically, the eigen and singular value decompositions at the core of
\code{tolerance_eigen()} and \code{tolerance_svd()} could be sped up and
or be more memory efficient with the usage of \pkg{RcppArmadillo}
\citep{eddelbuettel_rcpparmadillo_2014}, \pkg{RcppEigen}
\citep{bates_fast_2013}, or \pkg{RSpectra} \citep{qiu_rspectra_2019}.

\bibliography{GSVD_ms.bib}

\end{document}